\documentclass[11pt]{article}%
\usepackage{amsfonts}
\usepackage{amssymb}
\usepackage{amsmath}
\usepackage{amscd}
\usepackage{latexsym}
\usepackage[onehalfspacing]{setspace}
\usepackage{theorem}
\usepackage{natbib}
\usepackage[colorlinks=true,urlcolor=black,linkcolor=black,pdfpagemode="None",pdfstartview=FitH]{hyperref}
\usepackage{graphicx}
\usepackage{caption}
\providecommand{\U}[1]{\protect\rule{.1in}{.1in}}
\newtheorem{theorem}{Theorem}

\newcommand{\I}{{\rm 1\hspace*{-0.4ex}\rule{0.1ex}{1.52ex}\hspace*{0.2ex}}}

\hypersetup{pdfborder = {0 0 0},colorlinks=true,linkcolor=blue,citecolor=blue}
\renewcommand{\cite}{\citet*}
\begin{document}

\title{Average Density Estimators: Efficiency and Bootstrap Consistency\thanks{For
comments and suggestions, we are grateful to the Co-Editor, two referees, and
seminar participants at the Cowles Foundation conference celebrating Peter
Phillips' forty years at Yale, the 2018 LAMES conference, the 2019 CIREQ
Montreal Econometrics Conference, ITAM, University of North Carolina,
Princeton University, UC Santa Barbara, and Oxford University. Cattaneo
gratefully acknowledges financial support from the National Science Foundation
through grants SES-1459931 and SES-1947805, and Jansson gratefully
acknowledges financial support from the National Science Foundation through
grants SES-1459967 and SES-1947662, and the research support of CREATES
(funded by the Danish National Research Foundation under grant no. DNRF78).}}
\author{Matias D. Cattaneo\thanks{Department of Operations Research and Financial
Engineering, Princeton University.}
\and Michael Jansson\thanks{Department of Economics, University of California at
Berkeley and CREATES.}}
\maketitle

\begin{abstract}
This paper highlights a tension between semiparametric efficiency and
bootstrap consistency in the context of a canonical semiparametric estimation
problem, namely the problem of estimating the average density. It is shown
that although simple plug-in estimators suffer from bias problems preventing
them from achieving semiparametric efficiency under minimal smoothness
conditions, the nonparametric bootstrap automatically\ corrects for this bias
and that, as a result, these seemingly inferior estimators achieve bootstrap
consistency under minimal smoothness conditions. In contrast, several
\textquotedblleft debiased\textquotedblright\ estimators that achieve
semiparametric efficiency under minimal smoothness conditions do not achieve
bootstrap consistency under those same conditions.

\end{abstract}

\textit{Keywords:} Semiparametric estimation, efficiency, bootstrap consistency.

\textbf{JEL}: C14.%

\thispagestyle{empty}
\setcounter{page}{0}
\newpage

\section{Introduction\label{[Section] Introduction}}

Peter Phillips is a towering figure in econometrics. Among other things, his
pathbreaking work on nonstationary time series (e.g.,
\cite{Phillips_1987_ECMA} and \cite{Phillips-Perron_1988_Bio} in the case of
unit root autoregression and \cite{Phillips-Durlauf_1986_REStud} and
\cite{Phillips-Hansen_1990_REStud} in the case of cointegration) has
forcefully demonstrated that estimators can be useful without having limiting
distributions that are \textquotedblleft simple\textquotedblright. In this
paper, we show that a similar phenomenon occurs in a seemingly very different
setting, namely a canonical semiparametric estimation problem in a model with
$i.i.d.$ data.

The specific semiparametric estimation problem we consider is the problem of
estimating the average density of a continuously distributed random vector (of
which we have a random sample of observations). In that setting, a well known
apparent shortcoming of simple \textquotedblleft plug-in\textquotedblright%
\ estimators is that they have biases that are avoidable and potentially
non-negligible. In particular, the biases in question prevent the plug-in
estimators from achieving semiparametric efficiency under minimal smoothness
conditions. In recognition of this, several methods of \textquotedblleft
debiasing\textquotedblright\ have been proposed and have been found to be
successful insofar as they give rise to estimators that do achieve
semiparametric efficiency under minimal smoothness conditions. (The particular
examples given in this paper were obtained by applying and combining ideas
from \cite{Hall-Marron_1987_SPL}, \cite{Bickel-Ritov_1988_Sankhya}, and
\cite{Powell-Stock-Stoker_1989_ECMA}.)

Recognizing that construction of an estimator is often a means to the end of
conducting inference, a natural question is whether existing average density
estimators permit valid inference to be conducted under minimal smoothness
conditions. In this paper, we answer a specific version of the latter question
by investigating whether average density estimators achieve bootstrap
consistency under minimal smoothness conditions. Looking at estimators through
the lens of the bootstrap is of interest for several reasons, most notably
because one can answer questions motivated by inference considerations without
having to make additional (and potentially arbitrary) assumptions about the
behavior of standard errors (i.e., estimators of nuisance parameters). In
other words, because bootstrap consistency (or lack thereof) can be
interpreted as a property of an estimator, it has the potential to shed new
light on the relative merits of competing estimators. In this paper, we show
that average density estimation provides an example where this potential is realized.

To be specific, whereas several distinct approaches to debiasing achieve
semiparametric efficiency under minimal smoothness conditions, we find that
many of the estimators produced by these approaches fail to achieve bootstrap
consistency under minimal smoothness conditions. In contrast, in spite of
failing to achieve semiparametric efficiency under minimal smoothness
conditions, simple plug-in estimators achieve bootstrap consistency under
minimal smoothness conditions. In other words, we find that plug-in estimators
enjoy certain nontrivial advantages over some of their debiased counterparts.

The paper proceeds as follows. Section \ref{[Section] Setup} presents the
setup and introduces the formal questions we set out to answer. Studying the
most prominent average density estimators, Sections
\ref{[Section] Average Density Estimators: Efficiency} and
\ref{[Section] Average Density Estimators: Bootstrap Consistency} are
concerned with efficiency and bootstrap consistency, respectively. Alternative
estimators are analyzed in Section \ref{[Section] Alternative Estimators}.
Finally, Section \ref{[Section] Proofs} collects proofs of our main results.

\section{Setup\label{[Section] Setup}}

Suppose $X_{1},\ldots,X_{n}$ are $i.i.d.$ copies of a continuously distributed
random vector $X\in\mathbb{R}^{d}$ with an unknown density $f_{0}.$ Assuming
$f_{0}$ is square integrable, a widely studied estimand in this setting is%
\[
\theta_{0}=\mathbb{E}[f_{0}(X)],
\]
the average density. Influential work on estimating $\theta_{0}$ includes
\cite{Hall-Marron_1987_SPL}, \cite{Bickel-Ritov_1988_Sankhya}, and
\cite{Ritov-Bickel_1990_AoS}; see also \cite{Gine-Nickl_2008_Bernoulli} and
the references therein. In econometrics, estimators of $\theta_{0}$ are often
viewed as prototypical examples of two-step semiparametric estimators (in the
terminology of \cite{Newey-McFadden_1994_Handbook}) and therefore provide a
natural starting point when attempting to shed light on the properties of
two-step semiparametric estimators.

In what follows, we shall explore the extent to which certain prominent
estimators of $\theta_{0}$ enjoy one (or both) of two desirable properties.
The first of these properties is a very conventional one, namely
(semiparametric) efficiency. It is well known (e.g., \cite[Example
9.5.2]{Pfanzagl_1982_Book} and \cite{Ritov-Bickel_1990_AoS}) that if $f_{0}$
is bounded, then the efficient influence function $L_{0}$ is well-defined and
given by%
\[
L_{0}(x)=2\{f_{0}(x)-\theta_{0}\}.
\]
Accordingly, an estimator $\hat{\theta}_{n}=\hat{\theta}_{n}(X_{1}%
,\ldots,X_{n})$ of $\theta_{0}$ is said to be efficient if it satisfies%
\begin{equation}
\sqrt{n}(\hat{\theta}_{n}-\theta_{0})=\frac{1}{\sqrt{n}}\sum_{1\leq i\leq
n}L_{0}(X_{i})+o_{\mathbb{P}}(1).\label{Efficiency}%
\end{equation}

Our analysis will proceed under the following condition on the density.

\begin{description}
\item[Condition D] For some $s>d/4\ $with $2s\notin\mathbb{N},$ $f_{0}$ is
bounded and belongs to the Besov space $B_{2\infty}^{s}(\mathbb{R}^{d}).$
\end{description}

\noindent As alluded to earlier, the assumption that $f_{0}$ is bounded serves
the purpose of ensuring that%
\[
\sigma_{0}^{2}=\mathbb{V}[L_{0}(X)],
\]
the semiparametric variance bound implied by $\left(  \ref{Efficiency}\right)
,$ is well-defined and finite. As pointed out by
\cite{Bickel-Ritov_1988_Sankhya} and \cite{Ritov-Bickel_1990_AoS}, however,
some (additional) assumptions are required on the part of $f_{0}$ for
semiparametric efficiency to be achievable. For our purposes, it is convenient
and turns out to be sufficient to assume that $f_{0}$ is smooth in the sense
that it belongs to $B_{2\infty}^{s}(\mathbb{R}^{d}),$ as that assumption will
enable us to employ results from \cite{Gine-Nickl_2008_PTRF} when showing
asymptotic negligibility of certain remainder terms. In particular, and as
further discussed below, the magnitude \textquotedblleft
smoothing\textquotedblright\ bias of the kernel-based estimators under
consideration in this paper turns out to depend on $f_{0}$ through the
smoothness of the function $f_{0}^{\Delta}$ given by%
\[
f_{0}^{\Delta}(x)=\int_{\mathbb{R}^{d}}f_{0}(u)f_{0}(x+u)du.
\]
Condition D is convenient because it follows from \cite[Lemma 12]%
{Gine-Nickl_2008_PTRF} that $f_{0}^{\Delta}$ belongs to the H\"{o}lder space
$\mathbf{C}^{2s}(\mathbb{R}^{d})$ whenever $f_{0}$ is bounded and belongs to
$B_{2\infty}^{s}(\mathbb{R}^{d})$ with $2s\notin\mathbb{N}.$

The second property of interest is (nonparametric) bootstrap consistency. In
the setting of this paper, the most attractive definition of that property is
the following. Letting $X_{1,n}^{\ast},\ldots,X_{n,n}^{\ast}$ denote a random
sample from the empirical distribution of $X_{1},\ldots,X_{n}$ and letting
$\hat{\theta}_{n}^{\ast}=\hat{\theta}_{n}(X_{1,n}^{\ast},\ldots,X_{n,n}^{\ast
})$ denote the natural bootstrap analog of $\hat{\theta}_{n},$ the bootstrap
is said to be consistent if%
\begin{equation}
\sup_{t\in\mathbb{R}}\left\vert \mathbb{P}[\sqrt{n}(\hat{\theta}_{n}%
-\theta_{0})\leq t]-\mathbb{P}_{n}^{\ast}[\sqrt{n}(\hat{\theta}_{n}^{\ast
}-\hat{\theta}_{n})\leq t]\right\vert =o_{\mathbb{P}}%
(1),\label{Bootstrap consistency}%
\end{equation}
where $\mathbb{P}_{n}^{\ast}$ denotes a probability computed under the
bootstrap distribution conditional on the data.

To motivate interest in $\left(  \ref{Bootstrap consistency}\right)  ,$ recall
that the (nominal)\ level $1-\alpha$ bootstrap confidence interval for
$\theta_{0}$ based on the \textquotedblleft percentile
method\textquotedblright\ (in the terminology of \cite{vanderVaart_1998_Book})
is given by%
\[
\mathsf{CI}_{n,1-\alpha}^{\mathtt{P}}=\left[  \hat{\theta}_{n}-q_{n,1-\alpha
/2}^{\ast}~,~\hat{\theta}_{n}-q_{n,\alpha/2}^{\ast}\right]  ,\qquad
q_{n,a}^{\ast}=\inf\{q\in\mathbb{R}:\mathbb{P}_{n}^{\ast}[\hat{\theta}%
_{n}^{\ast}-\hat{\theta}_{n}\leq q]\geq a\}.
\]
This interval is said to be consistent if%
\begin{equation}
\lim_{n\rightarrow\infty}\mathbb{P}[\theta_{0}\in\mathsf{CI}_{n,1-\alpha
}^{\mathtt{P}}]=1-\alpha\label{Consistency of percentile interval}%
\end{equation}
and to be efficient if its end points satisfy%
\begin{equation}
\sqrt{n}(\hat{\theta}_{n}-q_{n,a}^{\ast}-\theta_{0})=\frac{1}{\sqrt{n}}%
\sum_{1\leq i\leq n}L_{0}(X_{i})-\Phi^{-1}(a)\sigma_{0}+o_{\mathbb{P}%
}(1),\qquad a\in\{\alpha/2,1-\alpha
/2\},\label{Efficiency of percentile interval}%
\end{equation}
where $\Phi(\cdot)$ is the standard normal cdf. In addition to being
\textquotedblleft heuristically necessary\textquotedblright, the bootstrap
consistency property $\left(  \ref{Bootstrap consistency}\right)  $ turns out
to be sufficient for $\left(  \ref{Consistency of percentile interval}\right)
$ and $\left(  \ref{Efficiency of percentile interval}\right)  $ in the cases
of interest in this paper. In turn, the property $\left(
\ref{Efficiency of percentile interval}\right)  $ implies (by the duality
between hypothesis tests and confidence intervals) that efficient two-sided
tests of simple hypotheses about $\theta_{0}$ can be based on $\mathsf{CI}%
_{n,1-\alpha}^{\mathtt{P}}$ whenever the interval is efficient. In other
words, the property $\left(  \ref{Bootstrap consistency}\right)  $ has strong
and obvious implications for inference and although those implications may
seem more important than bootstrap consistency per se, much of our subsequent
discussion of the bootstrap focuses on $\left(  \ref{Bootstrap consistency}%
\right)  $ for specificity and because that property seems more
\textquotedblleft fundamental\textquotedblright\ than $\left(
\ref{Consistency of percentile interval}\right)  $ and $\left(
\ref{Efficiency of percentile interval}\right)  $ in the sense that it is not
directly associated with a particular inference method.

At any rate, because the properties of $\hat{\theta}_{n}^{\ast}$ and
$\mathsf{CI}_{n,1-\alpha}^{\mathtt{P}}$ are governed solely by (the density
$f_{0}$ and) the functional form of $\hat{\theta}_{n},$ the properties
$\left(  \ref{Bootstrap consistency}\right)  ,$ $\left(
\ref{Consistency of percentile interval}\right)  ,$ and $\left(
\ref{Efficiency of percentile interval}\right)  $ can all be interpreted as
properties of the estimator $\hat{\theta}_{n}$ and one of the main purposes of
this paper is to explore the relationship between those properties and the
more familiar (efficiency) property $\left(  \ref{Efficiency}\right)  .$

The (nominal)\ level $1-\alpha$ bootstrap confidence interval for $\theta_{0}$
based on \textquotedblleft Efron's percentile method\textquotedblright\ (in
the terminology of \cite{vanderVaart_1998_Book}) is given by%
\[
\mathsf{CI}_{n,1-\alpha}^{\mathtt{E}}=\left[  \hat{\theta}_{n}+q_{n,\alpha
/2}^{\ast}~,~\hat{\theta}_{n}+q_{n,1-\alpha/2}^{\ast}\right]  .
\]
Suppose $\left(  \ref{Bootstrap consistency}\right)  $ holds. Then
$\mathsf{CI}_{n,1-\alpha}^{\mathtt{E}}$ is consistent if also $\left(
\ref{Efficiency}\right)  $ holds. On the other hand, and in contrast to
$\mathsf{CI}_{n,1-\alpha}^{\mathtt{P}},$ it turns out that in the cases of
interest in this paper the interval $\mathsf{CI}_{n,1-\alpha}^{\mathtt{E}}$ is
inconsistent when $\left(  \ref{Efficiency}\right)  $ fails. Partly for this
reason, we focus on intervals based on the percentile method.

Suppose $\left(  \ref{Efficiency}\right)  $ holds. Letting $\hat{\sigma}%
_{n}^{2}$ denote an estimator of $\sigma_{0}^{2},$ a natural (nominal)\ level
$1-\alpha$ confidence interval motivated by the distributional approximation
$\sqrt{n}(\hat{\theta}_{n}-\theta_{0})\overset{\cdot}{\sim}\mathcal{N}%
(0,\hat{\sigma}_{n}^{2})$ is the \textquotedblleft Normal\textquotedblright%
\ interval given by%
\[
\mathsf{CI}_{n,1-\alpha}^{\mathtt{N}}=\left[  \hat{\theta}_{n}-\Phi
^{-1}(1-\alpha/2)\hat{\sigma}_{n}/\sqrt{n}~,~\hat{\theta}_{n}-\Phi^{-1}%
(\alpha/2)\hat{\sigma}_{n}/\sqrt{n}\right]  .
\]
This interval is consistent if $\hat{\sigma}_{n}^{2}$ is consistent. The
bootstrap consistency property $\left(  \ref{Bootstrap consistency}\right)  $
is neither necessary nor sufficient for the \textquotedblleft bootstrap
variance consistency\textquotedblright\ property%
\begin{equation}
\hat{\sigma}_{n}^{2,\ast}=n\mathbb{V}[\hat{\theta}_{n}^{\ast}|X_{1}%
,\ldots,X_{n}]\rightarrow_{\mathbb{P}}\sigma_{0}^{2}%
.\label{Bootstrap variance consistency}%
\end{equation}
Following \cite{Bickel-Freedman_1981_AoS}, one way of ensuring that bootstrap
variance consistency is implied by bootstrap consistency is to employ the
Mallows metric $d_{2}$ when defining bootstrap consistency. The examples
studied herein have the feature that $(\ref{Bootstrap variance consistency})$
can hold even if $\left(  \ref{Bootstrap consistency}\right)  $ (and therefore
also convergence in the Mallows metric) fails. Partly for this reason it seems
more attractive (to us at least) to define bootstrap consistency as in
$\left(  \ref{Bootstrap consistency}\right)  ,$ hereby treating bootstrap
consistency and bootstrap variance consistency as distinct (i.e., non-nested) properties.

\section{Average Density Estimators:
Efficiency\label{[Section] Average Density Estimators: Efficiency}}

Our discussion of efficiency (or otherwise) of average density estimators
$\hat{\theta}_{n}$ will be based on the natural decomposition of the
estimation error $\hat{\theta}_{n}-\theta_{0}$ into its bias and
\textquotedblleft noise\textquotedblright\ components $\mathbb{E}[\hat{\theta
}_{n}]-\theta_{0}$ and $\hat{\theta}_{n}-\mathbb{E}[\hat{\theta}_{n}].\ $If
these components satisfy%
\begin{equation}
\sqrt{n}(\mathbb{E}[\hat{\theta}_{n}]-\theta_{0}%
)=o(1)\label{Efficiency: Bias condition}%
\end{equation}
and%
\begin{equation}
\sqrt{n}(\hat{\theta}_{n}-\mathbb{E}[\hat{\theta}_{n}])=\frac{1}{\sqrt{n}}%
\sum_{1\leq i\leq n}L_{0}(X_{i})+o_{\mathbb{P}}%
(1),\label{Efficiency: Noise condition}%
\end{equation}
respectively, then $\left(  \ref{Efficiency}\right)  $ holds. Moreover, if
$\left(  \ref{Efficiency: Noise condition}\right)  $ holds, then the
easy-to-interpret bias condition $\left(  \ref{Efficiency: Bias condition}%
\right)  $ is necessary and sufficient for $\left(  \ref{Efficiency}\right)
.$ The latter observation is particularly useful for our purposes, as it turns
out that the estimators of interest satisfy $\left(
\ref{Efficiency: Noise condition}\right)  $ under very mild conditions.

The simplest average density estimator is arguably the kernel-based
\textquotedblleft plug-in\textquotedblright\ estimator%
\[
\hat{\theta}_{n}^{\mathtt{AD}}=\frac{1}{n}\sum_{1\leq i\leq n}\hat{f}%
_{n}(X_{i}),
\]
where, for some kernel $K$ and some bandwidth $h_{n},$ $\hat{f}_{n}$ denotes
the kernel density estimator%
\[
\hat{f}_{n}\left(  x\right)  =\frac{1}{n}\sum_{1\leq j\leq n}K_{n}%
(x-X_{j}),\qquad K_{n}(x)=\frac{1}{h_{n}^{d}}K\left(  \frac{x}{h_{n}}\right)
.
\]

When developing results for $\hat{\theta}_{n}^{\mathtt{AD}}$ and other
estimators, we impose the following standard condition on the kernel, in which
$\Vert\cdot\Vert_{1}$ denotes the $\ell_{1}$-norm and $u^{l}$ is shorthand for
$u_{1}^{l_{1}}\cdots u_{d}^{l_{d}}$ when $u=(u_{1},\ldots,u_{d})^{\prime}%
\in\mathbb{R}^{d}$ and $l=(l_{1},\ldots,l_{d})^{\prime}\in\mathbb{Z}_{+}^{d}.$

\begin{description}
\item[Condition K] For some $P>d/2,$ $K$ is even and bounded with%
\[
\int_{\mathbb{R}^{d}}\left\vert K(u)\right\vert (1+\Vert u\Vert_{1}%
^{P})du<\infty
\]
and%
\[
\int_{\mathbb{R}^{d}}u^{l}K(u)du=\left\{
\begin{array}
[c]{ccl}%
1 &  & \text{if }l=0,\\
0 &  & \text{if }l\in\mathbb{Z}_{+}^{d}\text{ and }0<\Vert l\Vert_{1}<P.
\end{array}
\right.
\]

\end{description}

The constant $P$ in Condition K is the order of the kernel. Condition K
therefore implies that $K$ is a higher order kernel when $d\geq4.$ As usual,
we employ higher order kernels in order to ensure that the magnitude of the
smoothing bias of $\hat{f}_{n}$ is sufficiently small.

Under Conditions D and K, the density estimator $\hat{f}_{n}$ is consistent
(pointwise) provided the bandwidth satisfies

\begin{description}
\item[Condition B$^{-}$] As $n\rightarrow\infty,$ $h_{n}\rightarrow0$ and
$nh_{n}^{d}\rightarrow\infty.$
\end{description}

\noindent More importantly, Condition B$^{-}$ implies that the average density
estimator $\hat{\theta}_{n}^{\mathtt{AD}}$ satisfies $\left(
\ref{Efficiency: Noise condition}\right)  $ under Conditions D and
K.\footnote{Conversely, Condition B$^{-}$ is minimal in the sense that the
methods of \cite{Cattaneo-Crump-Jansson_2014_ET_SmallBW} can be used to show
that $\left(  \ref{Efficiency: Noise condition}\right)  $ can fail if
Condition B$^{-}$ is violated.} As a consequence, under Conditions D, K, and
B$^{-},$ the estimator $\hat{\theta}_{n}^{\mathtt{AD}}$ is efficient if and
only if it satisfies the bias condition $\left(
\ref{Efficiency: Bias condition}\right)  .$

Using the representation $\theta_{0}=f_{0}^{\Delta}(0),\ $the bias of
$\hat{\theta}_{n}^{\mathtt{AD}}$ can be shown to admit the approximation%
\begin{equation}
\mathbb{E}[\hat{\theta}_{n}^{\mathtt{AD}}]-\theta_{0}\approx\frac{K(0)}%
{nh_{n}^{d}}+\int_{\mathbb{R}^{d}}K(t)[f_{0}^{\Delta}(h_{n}t)-f_{0}^{\Delta
}(0)]dt,\label{Plug-In AD estimator: Bias}%
\end{equation}
where the approximation error is of order $n^{-1},$ the first term is a
\textquotedblleft leave in\textquotedblright\ bias term (in the terminology of
\cite{Cattaneo-Crump-Jansson_2013_JASA}), and the second term is a smoothing
bias term. As previosly mentioned, the function $f_{0}^{\Delta}$ belongs to
the H\"{o}lder space $\mathbf{C}^{2s}(\mathbb{R}^{d})$ under Condition D.
Using this fact, it follows from a routine calculation (e.g.,
\cite[Proposition 1.2]{Tsybakov_2009_Book}) that if Conditions D and K are
satisfied and if $h_{n}\rightarrow0,$ then%
\[
\int_{\mathbb{R}^{d}}K(t)[f_{0}^{\Delta}(h_{n}t)-f_{0}^{\Delta}(0)]dt=O(h_{n}%
^{2S}),\qquad S=\min(P/2,s).
\]

As a consequence, under Conditions D and K the estimator $\hat{\theta}%
_{n}^{\mathtt{AD}}$ is efficient provided Condition B$^{-}$ is strengthened to

\begin{description}
\item[Condition B$^{+}$] As $n\rightarrow\infty,$ $nh_{n}^{4S}\rightarrow0 $
and $nh_{n}^{2d}\rightarrow\infty.$
\end{description}

\noindent Existence of a bandwidth sequence satisfying Condition B$^{+}$
requires that the parameter $s$ governing the smoothness of $f_{0}$ satisfies
$s>d/2,$ a stronger condition than the (minimal) condition $s>d/4$ included in
Condition D.

This shortcoming of $\hat{\theta}_{n}^{\mathtt{AD}}$ is attributable to its
leave in bias, as it is the presence of the leave in bias that requires a
strengthening of the lower bound on the bandwidth from $nh_{n}^{d}%
\rightarrow\infty$ to $nh_{n}^{2d}\rightarrow\infty.$ Of course, the leave in
bias of $\hat{\theta}_{n}^{\mathtt{AD}}$ is easily avoidable. One option is to
employ a kernel satisfying $K(0)=0.$ Recognizing that all standard kernels
have $K(0)\neq0,$ a more natural option is to use the \textquotedblleft
bias-corrected\textquotedblright\ version of $\hat{\theta}_{n}^{\mathtt{AD}}$
given by%
\[
\hat{\theta}_{n}^{\mathtt{AD-BC}}=\hat{\theta}_{n}^{\mathtt{AD}}-\frac
{K(0)}{nh_{n}^{d}}.
\]
By construction, the bias of this estimator satisfies%
\[
\mathbb{E}[\hat{\theta}_{n}^{\mathtt{AD-BC}}]-\theta_{0}\approx\int%
_{\mathbb{R}^{d}}K(t)[f_{0}^{\Delta}(h_{n}t)-f_{0}^{\Delta}(0)]dt=O(h_{n}%
^{2S}),
\]
so under Conditions D and K the bias condition $\left(
\ref{Efficiency: Bias condition}\right)  $ is satisfied by $\hat{\theta}%
_{n}^{\mathtt{AD-BC}}$ provided $nh_{n}^{4S}\rightarrow0,$ implying in turn
that $\hat{\theta}_{n}^{\mathtt{AD-BC}}$ is asymptotically efficient under
Conditions D and K provided the bandwidth satisfies the following condition,
which requires no additional smoothness (as measured by the value of $s$)
relative to Condition D.

\begin{description}
\item[Condition B] As $n\rightarrow\infty,$ $nh_{n}^{4S}\rightarrow0$ and
$nh_{n}^{d}\rightarrow\infty.$
\end{description}

The leave in bias of $\hat{\theta}_{n}^{\mathtt{AD}}$ is proportional to
$1/(nh_{n}^{d}).$ Equipped with only that knowledge, the method of generalized
jackknifing constructs a debiased version of $\hat{\theta}_{n}^{\mathtt{AD}}$
as a weighted sum of two (or more) versions of $\hat{\theta}_{n}^{\mathtt{AD}%
}$ implemented using different values of the bandwidth, where the weights are
judiciously chosen to remove the leave in bias. To give the simplest example,
let $\hat{\theta}_{n}^{\mathtt{AD}}(h)$ denote the version of $\hat{\theta
}_{n}^{\mathtt{AD}}$ associated with the bandwidth $h.$ Then, for any $c\neq1$
the \textquotedblleft generalized jackknife\textquotedblright\ version of
$\hat{\theta}_{n}^{\mathtt{AD}}$ obtained by combining $\hat{\theta}%
_{n}^{\mathtt{AD}}=\hat{\theta}_{n}^{\mathtt{AD}}(h_{n})$ and $\hat{\theta
}_{n}^{\mathtt{AD}}(ch_{n})$ is given by%
\[
\hat{\theta}_{n}^{\mathtt{AD-GJ}}=\frac{1}{1-c^{d}}\hat{\theta}_{n}%
^{\mathtt{AD}}-\frac{c^{d}}{1-c^{d}}\hat{\theta}_{n}^{\mathtt{AD}}(ch_{n}).
\]
Like $\hat{\theta}_{n}^{\mathtt{AD}},$ the estimator $\hat{\theta}%
_{n}^{\mathtt{AD-GJ}}$ satisfies $\left(  \ref{Efficiency: Noise condition}%
\right)  $ under Conditions D, K, and B$^{-}$. Moreover, because%
\[
\mathbb{E}[\hat{\theta}_{n}^{\mathtt{AD}}(ch_{n})]-\theta_{0}\approx\frac
{1}{c^{d}}\frac{K(0)}{nh_{n}^{d}}+\int_{\mathbb{R}^{d}}K(t)[f_{0}^{\Delta
}(ch_{n}t)-f_{0}^{\Delta}(0)]dt,
\]
the bias condition $\left(  \ref{Efficiency: Bias condition}\right)  $ is
satisfied by $\hat{\theta}_{n}^{\mathtt{AD-GJ}}$ under Condition B.

Finally, as its name suggests, the leave in bias can also be avoided by
employing \textquotedblleft leave out\textquotedblright\ estimators of
$f_{0}.$ A generic average density estimator based on leave out density
estimators is of the form%
\[
\hat{\theta}_{n}^{\mathtt{AD-LO}}=\frac{1}{n}\sum_{1\leq i\leq n}\hat{f}%
_{i,n}^{\mathtt{LO}}(X_{i}),
\]
where $\hat{f}_{i,n}^{\mathtt{LO}}$ is a kernel density estimator constructed
using observations belonging to a set that does not include $X_{i}.$ Relative
to $\hat{\theta}_{n}^{\mathtt{AD-BC}}$ and $_{n}^{\mathtt{AD-GJ}},$ an
attractive feature of $\hat{\theta}_{n}^{\mathtt{AD-LO}}$ is that it can be
constructed without knowledge of the functional form of the leave-in bias. For
concreteness, we shall develop results for $\hat{\theta}_{n}^{\mathtt{AD-LO}}$
only in the (leading) special case where the sample $X_{1},\ldots,X_{n}$ is
partitioned into $B_{n}\in\{2,\ldots,n\}$ disjoint blocks of (approximately)
equal size and $\hat{f}_{i,n}^{\mathtt{LO}}$ is constructed using observations
from all blocks except the one to which the $i $th observation belongs. To be
specific, we assume that $\hat{f}_{i,n}^{\mathtt{LO}}$ is of the form%
\[
\hat{f}_{i,n}^{\mathtt{LO}}(x)=\sum_{1\leq j\leq n}w_{ij,n}K_{n}%
(x-X_{j}),\qquad w_{ij,n}=\frac{%
\I
(\left\lceil iB_{n}/n\right\rceil \neq\left\lceil jB_{n}/n\right\rceil )}%
{\sum_{1\leq k\leq n}%
\I
(\left\lceil iB_{n}/n\right\rceil \neq\left\lceil kB_{n}/n\right\rceil )}.
\]

When $B_{n}=n,$ $\hat{f}_{i,n}^{\mathtt{LO}}$ is the $i$th \textquotedblleft
leave-one-out\textquotedblright\ estimator of $f_{0}$ and the estimator
$\hat{\theta}_{n}^{\mathtt{AD-LO}}$ reduces to the estimator introduced in
\cite{Hall-Marron_1987_SPL} and further studied by
\cite{Gine-Nickl_2008_Bernoulli} (among many others). At the opposite extreme,
when $B_{n}$ is kept fixed, the estimator $\hat{\theta}_{n}^{\mathtt{AD-LO}}$
is a \textquotedblleft cross-fit\textquotedblright\ estimator (using an
$B_{n}$-fold non-random partition of $\{1,\ldots,n\}$) in the terminology of
\cite{Newey-Robins_2018_CrossFitting}.

Regardless of the choice of $B_{n},$ under Conditions D, K, and B$^{-},$ the
estimator $\hat{\theta}_{n}^{\mathtt{AD-LO}}$ is similar to $\hat{\theta}%
_{n}^{\mathtt{AD-BC}}$ and $\hat{\theta}_{n}^{\mathtt{AD-GJ}}$ insofar as it
satisfies $\left(  \ref{Efficiency: Noise condition}\right)  $ and has%
\[
\mathbb{E}[\hat{\theta}_{n}^{\mathtt{AD-LO}}]-\theta_{0}\approx\int%
_{\mathbb{R}^{d}}K(t)[f_{0}^{\Delta}(h_{n}t)-f_{0}^{\Delta}(0)]dt=O(h_{n}%
^{2S}),
\]
implying in particular that $\hat{\theta}_{n}^{\mathtt{AD-LO}}$ is
asymptotically efficient under Conditions D, K and B.

The following result collects and summarizes the main findings of this section.

\begin{theorem}
\label{[Theorem] Efficiency}Suppose Conditions D, K, and B are satisfied. Then
$\hat{\theta}_{n}^{\mathtt{AD-BC}},$ $\hat{\theta}_{n}^{\mathtt{AD-GJ}}, $ and
$\hat{\theta}_{n}^{\mathtt{AD-LO}}$ satisfy $\left(  \ref{Efficiency}\right)
.$ If Condition B is strengthened to Condition B$^{+}$, then $\hat{\theta}%
_{n}^{\mathtt{AD}}$ satisfies $\left(  \ref{Efficiency}\right)  .$
\end{theorem}

\noindent\textbf{Remark.} Because $\hat{\theta}_{n}^{\mathtt{AD}}$ is linear
functional of $\hat{f}_{n},$ the generalized jackknife estimator $\hat{\theta
}_{n}^{\mathtt{AD-GJ}}$ can be interpreted as a version of the plug-in
estimator $\hat{\theta}_{n}^{\mathtt{AD}}$ based on a modified kernel:
Defining%
\[
K^{\mathtt{GJ}}(x)=\frac{1}{1-c^{d}}\left[  K(x)-K\left(  \frac{x}{c}\right)
\right]  ,
\]
we have%
\[
\hat{\theta}_{n}^{\mathtt{AD-GJ}}=\frac{1}{n}\sum_{1\leq i\leq n}\hat{f}%
_{n}^{\mathtt{GJ}}(X_{i}),
\]
where%
\[
\hat{f}_{n}^{\mathtt{AD-GJ}}\left(  x\right)  =\frac{1}{n}\sum_{1\leq j\leq
n}K_{n}^{\mathtt{GJ}}(x-X_{j}),\qquad K_{n}^{\mathtt{GJ}}(x)=\frac{1}%
{h_{n}^{d}}K^{\mathtt{GJ}}\left(  \frac{x}{h_{n}}\right)  .
\]
The modified kernel satisfies $K^{\mathtt{GJ}}(0)=0,$ so this interpretation
provides an explanation of the fact that $\hat{\theta}_{n}^{\mathtt{AD-GJ}}$
satisfies $\left(  \ref{Efficiency: Bias condition}\right)  $ under Condition
B. A similar interpretation is not available for generalized jackknife
versions of estimators that are nonlinear functionals of $\hat{f}_{n};$
examples of such estimators are given by $\hat{\theta}_{n}^{\mathtt{ISD-GJ}}$
and $\hat{\theta}_{n}^{\mathtt{LR-GJ}}$ studied in Section
\ref{[Section] Alternative Estimators}.

\section{Average Density Estimators: Bootstrap
Consistency\label{[Section] Average Density Estimators: Bootstrap Consistency}%
}

Letting $X_{1,n}^{\ast},\ldots,X_{n,n}^{\ast}$ denote a random sample from the
empirical distribution of $X_{1},\ldots,X_{n},$ the natural bootstrap analogs
of the estimators studied in the previous section are given by%
\[
\hat{\theta}_{n}^{\mathtt{AD,}\ast}=\frac{1}{n}\sum_{1\leq i\leq n}\hat{f}%
_{n}^{\ast}(X_{i,n}^{\ast}),\qquad\hat{f}_{n}^{\ast}\left(  x\right)
=\frac{1}{n}\sum_{1\leq j\leq n}K_{n}(x-X_{j,n}^{\ast}),
\]%
\[
\hat{\theta}_{n}^{\mathtt{AD-BC,}\ast}=\hat{\theta}_{n}^{\mathtt{AD,}\ast
}-\frac{K(0)}{nh_{n}^{d}},
\]%
\[
\hat{\theta}_{n}^{\mathtt{AD-GJ,}\ast}=\frac{1}{1-c^{d}}\hat{\theta}%
_{n}^{\mathtt{AD,}\ast}-\frac{c^{d}}{1-c^{d}}\hat{\theta}_{n}^{\mathtt{AD,}%
\ast}(ch_{n}),
\]
and%
\[
\hat{\theta}_{n}^{\mathtt{AD-LO,}\ast}=\frac{1}{n}\sum_{1\leq i\leq n}\hat
{f}_{i,n}^{\mathtt{LO,}\ast}(X_{i,n}^{\ast}),\qquad\hat{f}_{i,n}%
^{\mathtt{LO,}\ast}(x)=\sum_{1\leq j\leq n}w_{ij,n}K_{n}(x-X_{j,n}^{\ast}),
\]
respectively, where $\hat{\theta}_{n}^{\mathtt{AD,}\ast}(ch_{n})$ denotes the
version of $\hat{\theta}_{n}^{\mathtt{AD,}\ast}$ associated with the bandwidth
$ch_{n}.$ The main goal of this section is to explore the extent to which
these estimators enjoy the bootstrap consistency property $\left(
\ref{Bootstrap consistency}\right)  $ under Conditions D, K, and B.

If $\hat{\theta}_{n}$ is efficient in the sense that it satisfies $\left(
\ref{Efficiency}\right)  ,$ then $\sqrt{n}(\hat{\theta}_{n}-\theta
_{0})\rightsquigarrow\mathcal{N}(0,\sigma_{0}^{2}),$ implying in particular
that the bootstrap consistency property $\left(  \ref{Bootstrap consistency}%
\right)  $ admits the following characterization:%
\begin{equation}
\sqrt{n}(\hat{\theta}_{n}^{\ast}-\hat{\theta}_{n})\rightsquigarrow
_{\mathbb{P}}\mathcal{N}(0,\sigma_{0}^{2}%
),\label{Bootstrap consistency: Characterization under efficiency}%
\end{equation}
where $\rightsquigarrow_{\mathbb{P}}$ denotes conditional weak convergence in probability.

Similarly to the analysis of the previous section, it seems natural to base
verification of $\left(
\ref{Bootstrap consistency: Characterization under efficiency}\right)  $ on a
decomposition of the bootstrap estimation error $\hat{\theta}_{n}^{\ast}%
-\hat{\theta}_{n}$ into its bias and noise\ components $\mathbb{E}_{n}^{\ast
}[\hat{\theta}_{n}^{\ast}]-\hat{\theta}_{n} $ and $\hat{\theta}_{n}^{\ast
}-\mathbb{E}_{n}^{\ast}[\hat{\theta}_{n}^{\ast}],$ where $\mathbb{E}_{n}%
^{\ast}[\cdot]=\mathbb{E}[\cdot|X_{1},\ldots,X_{n}].$ The resulting sufficient
condition for $\left(
\ref{Bootstrap consistency: Characterization under efficiency}\right)  $ is
given by the pair%
\begin{equation}
\sqrt{n}(\mathbb{E}_{n}^{\ast}[\hat{\theta}_{n}^{\ast}]-\hat{\theta}%
_{n})=o_{\mathbb{P}}%
(1)\label{Bootstrap consistency: Bias condition under efficiency}%
\end{equation}
and%
\begin{equation}
\sqrt{n}(\hat{\theta}_{n}^{\ast}-\mathbb{E}_{n}^{\ast}[\hat{\theta}_{n}^{\ast
}])\rightsquigarrow_{\mathbb{P}}\mathcal{N}(0,\sigma_{0}^{2}%
),\label{Bootstrap consistency: Noise condition}%
\end{equation}
where $\left(  \ref{Bootstrap consistency: Bias condition under efficiency}%
\right)  $ is the natural bootstrap analog of $\left(
\ref{Efficiency: Bias condition}\right)  ,$ $\left(
\ref{Bootstrap consistency: Noise condition}\right)  $ is a bootstrap version
of the main distributional implication of $\left(
\ref{Efficiency: Noise condition}\right)  ,$ and where $\left(
\ref{Bootstrap consistency: Bias condition under efficiency}\right)  $ is
necessary and sufficient for $\left(
\ref{Bootstrap consistency: Characterization under efficiency}\right)  $ when
$\left(  \ref{Bootstrap consistency: Noise condition}\right)  $ holds.

In perfect analogy with $\left(  \ref{Efficiency: Noise condition}\right)  ,$
it turns out that $\left(  \ref{Bootstrap consistency: Noise condition}%
\right)  $ holds under very mild bandwidth conditions. Indeed, under
Conditions D and K, the estimators $\hat{\theta}_{n}^{\mathtt{AD,}\ast},$
$\hat{\theta}_{n}^{\mathtt{AD-BC,}\ast},$ $\hat{\theta}_{n}^{\mathtt{AD-GJ,}%
\ast},$ and $\hat{\theta}_{n}^{\mathtt{AD-LO,}\ast}$ all satisfy $\left(
\ref{Bootstrap consistency: Noise condition}\right)  $ whenever Condition
B$^{-}$ holds.\footnote{Conversely, Condition B$^{-}$ is minimal in the sense
that the methods of \cite{Cattaneo-Crump-Jansson_2014_ET_Bootstrap} can be
used to show that $\left(  \ref{Bootstrap consistency: Noise condition}%
\right)  $ can fail if Condition B$^{-}$ is violated.} As a consequence, the
question once again becomes whether the estimators have biases that are
sufficiently small. Under Conditions D, K, and B$^{-}$, the bootstrap bias of
$\hat{\theta}_{n}^{\mathtt{AD,}\ast}$ satisfies%
\begin{equation}
\mathbb{E}_{n}^{\ast}[\hat{\theta}_{n}^{\mathtt{AD,}\ast}]-\hat{\theta}%
_{n}^{\mathtt{AD}}=\frac{K(0)}{nh_{n}^{d}}-\frac{1}{n}\hat{\theta}%
_{n}^{\mathtt{AD}}=\frac{K(0)}{nh_{n}^{d}}+O_{\mathbb{P}}(n^{-1}%
).\label{Plug-In AD estimator: Bootstrap bias}%
\end{equation}
Therefore, the bias condition $\left(
\ref{Bootstrap consistency: Bias condition under efficiency}\right)  $ is
satisfied by $\hat{\theta}_{n}^{\mathtt{AD,}\ast}$ provided $nh_{n}%
^{2d}\rightarrow\infty.$ In other words, $\hat{\theta}_{n}^{\mathtt{AD,}\ast}$
satisfies $\left(  \ref{Bootstrap consistency}\right)  $ (and therefore also
$\left(  \ref{Consistency of percentile interval}\right)  $ and $\left(
\ref{Efficiency of percentile interval}\right)  $) under Conditions D, K, and
B$^{+}$.

More surprisingly, perhaps, although the estimator $\hat{\theta}%
_{n}^{\mathtt{AD-BC}}$ is efficient under Conditions D, K, and B, stronger
conditions are required for its bootstrap analog $\hat{\theta}_{n}%
^{\mathtt{AD-BC,}\ast}$ to satisfy $\left(  \ref{Bootstrap consistency}%
\right)  .$ This is so because%
\begin{equation}
\mathbb{E}_{n}^{\ast}[\hat{\theta}_{n}^{\mathtt{AD-BC,}\ast}]-\hat{\theta}%
_{n}^{\mathtt{AD-BC}}=\mathbb{E}_{n}^{\ast}[\hat{\theta}_{n}^{\mathtt{AD,}%
\ast}]-\hat{\theta}_{n}^{\mathtt{AD}}=\frac{K(0)}{nh_{n}^{d}}+O_{\mathbb{P}%
}(n^{-1})\label{Bias-Corrected AD estimator: Bootstrap bias}%
\end{equation}
under Conditions D, K, and B. A similar remark applies to $\hat{\theta}%
_{n}^{\mathtt{AD-LO}},$ as its bootstrap analog satisfies%
\[
\mathbb{E}_{n}^{\ast}[\hat{\theta}_{n}^{\mathtt{AD-LO,}\ast}]-\hat{\theta}%
_{n}^{\mathtt{AD-LO}}=\hat{\theta}_{n}^{\mathtt{AD}}-\hat{\theta}%
_{n}^{\mathtt{AD-LO}}=\frac{K(0)}{nh_{n}^{d}}+o_{\mathbb{P}}(n^{-1/2})
\]
under Conditions D, K, and B.

On the other hand, because%
\[
\mathbb{E}_{n}^{\ast}[\hat{\theta}_{n}^{\mathtt{AD,}\ast}(ch_{n})]-\hat
{\theta}_{n}^{\mathtt{AD}}(ch_{n})=\frac{1}{c^{d}}\frac{K(0)}{nh_{n}^{d}%
}-\frac{1}{n}\hat{\theta}_{n}^{\mathtt{AD}}(ch_{n})=\frac{1}{c^{d}}\frac
{K(0)}{nh_{n}^{d}}+O_{\mathbb{P}}(n^{-1}),
\]
the bootstrap analog of $\hat{\theta}_{n}^{\mathtt{AD-GJ}}$ satisfies%
\[
\mathbb{E}_{n}^{\ast}[\hat{\theta}_{n}^{\mathtt{AD-GJ,}\ast}]-\hat{\theta}%
_{n}^{\mathtt{AD-GJ}}=O_{\mathbb{P}}(n^{-1}),
\]
so this estimator satisfies $\left(  \ref{Bootstrap consistency}\right)  $
under Conditions D, K, and B.

It turns out that $\hat{\theta}_{n}^{\mathtt{AD,}\ast}$\ satisfies $\left(
\ref{Bootstrap consistency}\right)  ,$ $\left(
\ref{Consistency of percentile interval}\right)  ,$ and $\left(
\ref{Efficiency of percentile interval}\right)  $ under conditions that are
weaker than the conditions under which $\hat{\theta}_{n}^{\mathtt{AD}}$ is
efficient. In generic notation, suppose the estimators $\hat{\theta}_{n}$ and
$\hat{\theta}_{n}^{\ast}$ satisfy $\left(  \ref{Efficiency: Noise condition}%
\right)  $ and $\left(  \ref{Bootstrap consistency: Noise condition}\right)
,$ respectively. Then $\left(  \ref{Bootstrap consistency}\right)  $ is still
sufficient for $\left(  \ref{Consistency of percentile interval}\right)  ,$
and $\left(  \ref{Efficiency of percentile interval}\right)  $ to hold.
Moreover, as also observed by \cite{Cattaneo-Jansson_2018_ECMA}, the bootstrap
consistency condition $\left(  \ref{Bootstrap consistency}\right)  $ itself is
satisfied under the following generalization of the bias conditions $\left(
\ref{Efficiency: Bias condition}\right)  $ and $\left(
\ref{Bootstrap consistency: Bias condition under efficiency}\right)  :$%
\begin{equation}
\sqrt{n}(\mathbb{E}_{n}^{\ast}[\hat{\theta}_{n}^{\ast}]-\hat{\theta}%
_{n})=\sqrt{n}(\mathbb{E}[\hat{\theta}_{n}]-\theta_{0})+o_{\mathbb{P}%
}(1).\label{Bootstrap consistency: Bias condition}%
\end{equation}
Now, as discussed above, the estimators $\hat{\theta}_{n}^{\mathtt{AD}}$ and
$\hat{\theta}_{n}^{\mathtt{AD,}\ast}$ satisfy $\left(
\ref{Efficiency: Noise condition}\right)  $ and $\left(
\ref{Bootstrap consistency: Noise condition}\right)  ,$ respectively, under
Conditions D, K, and B. Under the same conditions, it follows from $\left(
\ref{Plug-In AD estimator: Bias}\right)  $ and $\left(
\ref{Plug-In AD estimator: Bootstrap bias}\right)  $\ that $\left(
\ref{Bootstrap consistency: Bias condition}\right)  $ is satisfied.

The following result collects and summarizes the main findings of this section.

\begin{theorem}
\label{[Theorem] Bootstrap consistency}Suppose Conditions D, K, and B are
satisfied. Then $\hat{\theta}_{n}^{\mathtt{AD,}\ast}$ and $\hat{\theta}%
_{n}^{\mathtt{AD-GJ,}\ast}$ satisfy $\left(  \ref{Bootstrap consistency}%
\right)  .$ If Condition B is strengthened to Condition B$^{+}$, then
$\hat{\theta}_{n}^{\mathtt{AD-BC,}\ast}$ and $\hat{\theta}_{n}%
^{\mathtt{AD-LO,}\ast}$ satisfy $\left(  \ref{Bootstrap consistency}\right)
.$
\end{theorem}

Comparing Theorems \ref{[Theorem] Efficiency} and
\ref{[Theorem] Bootstrap consistency}, we see that efficiency is neither
necessary nor sufficient for bootstrap consistency. In fact, the results
indicate that there can be a tension between efficiency and bootstrap
consistency in semiparametric settings. What seems most noteworthy to us is
that whereas \textquotedblleft debiased\textquotedblright\ estimators such as
$\hat{\theta}_{n}^{\mathtt{AD-BC}}$ and $\hat{\theta}_{n}^{\mathtt{AD-LO}}$
may appear to be superior to the simple plug-in estimator $\hat{\theta}%
_{n}^{\mathtt{AD}}$ insofar as they achieve efficiency under weaker (indeed,
minimal) conditions, the ranking gets reversed when the estimators are looked
at through the lens of the bootstrap. As pointed out by
\cite{Chen-Linton-vanKeilegom_2003_ECMA} and \cite{Cheng-Huang_2010_AoS},
bootstrap-based inference is particularly attractive in semiparametric
settings. The results above demonstrate by example that efficiency-based
rankings of estimators can be quite misleading\ in cases where construction of
an estimator is simply a means to the end of conducting bootstrap-based inference.

As conjectured by the co-editor, the estimators $\hat{\theta}_{n}%
^{\mathtt{AD}},$ $\hat{\theta}_{n}^{\mathtt{AD-BC}},$ $\hat{\theta}%
_{n}^{\mathtt{AD-GJ}}, $ and $\hat{\theta}_{n}^{\mathtt{AD-LO}}$ can all be
shown to satisfy the bootstrap variance consistency property
$(\ref{Bootstrap variance consistency})$ under Conditions D, K, and B. The
estimators $\hat{\theta}_{n}^{\mathtt{AD-BC}},$ $\hat{\theta}_{n}%
^{\mathtt{AD-GJ}},$ and $\hat{\theta}_{n}^{\mathtt{AD-LO}}$ therefore enjoy
the property that the intervals $\mathsf{CI}_{n,1-\alpha}^{\mathtt{N}}$ based
on the bootstrap variance estimator are consistent (indeed, efficient) under
Conditions D, K, and B.

An important source of the bootstrap consistency result for $\hat{\theta}%
_{n}^{\mathtt{AD}}$ is the ability of the bootstrap to automatically perform a
bias correction when approximating the distribution of $\hat{\theta}%
_{n}^{\mathtt{AD}}-\theta_{0}.$ The same mechanism can be exploited for
estimation purposes: Setting $\alpha=1,$ the interval $\mathsf{CI}%
_{n,1-\alpha}^{\mathtt{P}}$ becomes a singleton and can therefore be
interpreted as a bootstrap-based estimator of $\theta_{0}.$ As a by-product of
our results about $\hat{\theta}_{n}^{\mathtt{AD}},$ it can be shown that the
resulting estimator%
\[
2\hat{\theta}_{n}^{\mathtt{AD}}-\inf\{q\in\mathbb{R}:\mathbb{P}_{n}^{\ast
}[\hat{\theta}_{n}^{\mathtt{AD,}\ast}\leq q]\geq1/2\},
\]
is efficient under Conditions D, K, and B.

The bootstrap analog of $\hat{\theta}_{n}^{\mathtt{AD}}$ employs a density
estimator $\hat{f}_{n}^{\ast}$ that uses the same bandwidth $h_{n}$ as is used
when constructing $\hat{f}_{n}.$ Doing so is important for the purposes of
obtaining the bootstrap consistency result for $\hat{\theta}_{n}^{\mathtt{AD}%
}.$ Indeed, if $\hat{f}_{n}^{\ast}$ were defined using a possibly different
bandwidth $h_{n}^{\ast}$ (say), then the bootstrap consistency result under
Condition B can fail unless $h_{n}^{\ast}/h_{n}\rightarrow_{\mathbb{P}}1.$ On
the other hand, the flavor of the bootstrap results about $\hat{\theta}%
_{n}^{\mathtt{AD-BC}},$ $\hat{\theta}_{n}^{\mathtt{AD-LO}},$ and $\hat{\theta
}_{n}^{\mathtt{AD-GJ}}$ does not change if a different bandwidth is used when
defining their bootstrap analogs.

In light of Theorem \ref{[Theorem] Bootstrap consistency} it is of interest to
construct bootstrap-based approximations to the distributions of $\hat{\theta
}_{n}^{\mathtt{AD-BC}}$ and $\hat{\theta}_{n}^{\mathtt{AD-LO}}$ that are
consistent under Conditions D, K, and B. In generic notation, suppose
$\hat{\theta}_{n}=\hat{\theta}_{n}(X_{1},\ldots,X_{n})$ is the estimator whose
distribution we seek to approximate. One option is to find an estimator
$\tilde{\theta}_{n}=\tilde{\theta}_{n}(X_{1},\ldots,X_{n})$ (say) whose
natural bootstrap analog $\tilde{\theta}_{n}^{\ast}=\tilde{\theta}_{n}%
(X_{1,n}^{\ast},\ldots,X_{n,n}^{\ast})$ satisfies%
\begin{equation}
\sup_{t\in\mathbb{R}}\left\vert \mathbb{P}[\sqrt{n}(\hat{\theta}_{n}%
-\theta_{0})\leq t]-\mathbb{P}_{n}^{\ast}[\sqrt{n}(\tilde{\theta}_{n}^{\ast
}-\hat{\theta}_{n})\leq t]\right\vert =o_{\mathbb{P}}%
(1).\label{Bootstrap consistency: Alternative estimator}%
\end{equation}
As we shall see, both $\hat{\theta}_{n}^{\mathtt{AD-BC}}$ and $\hat{\theta
}_{n}^{\mathtt{AD-LO}}$ lend themselves well to a construction of this type.
Nevertheless, in some circumstances it may be equally (if not more) attractive
to achieve consistency by finding a bootstrap probability measure
$\mathbb{P}_{n}^{\star}$ (say) governing the distribution of $X_{1,n}^{\star
},\ldots,X_{n,n}^{\star}$ such that $\hat{\theta}_{n}^{\star}=\hat{\theta}%
_{n}(X_{1,n}^{\star},\ldots,X_{n,n}^{\star})$ satisfies%

\begin{equation}
\sup_{t\in\mathbb{R}}\left\vert \mathbb{P}[\sqrt{n}(\hat{\theta}_{n}%
-\theta_{0})\leq t]-\mathbb{P}_{n}^{\star}[\sqrt{n}(\hat{\theta}_{n}^{\star
}-\hat{\theta}_{n})\leq t]\right\vert =o_{\mathbb{P}}%
(1).\label{Bootstrap consistency: Alternative measure}%
\end{equation}
A construction of this type turns out to be useful in the case of the
cross-fit version of $\hat{\theta}_{n}^{\mathtt{AD-LO}}.$

First, consider the problem of approximating the distribution of $\hat{\theta
}_{n}^{\mathtt{AD-BC}}.$ It follows from $\left(
\ref{Bias-Corrected AD estimator: Bootstrap bias}\right)  $ that a
bias-corrected version of $\hat{\theta}_{n}^{\mathtt{AD-BC,}\ast}$ is given by%
\[
\tilde{\theta}_{n}^{\mathtt{AD-BC,}\ast}=\hat{\theta}_{n}^{\mathtt{AD-BC,}%
\ast}-\frac{K(0)}{nh_{n}^{d}}.
\]
Rather than showing $\left(
\ref{Bootstrap consistency: Alternative estimator}\right)  $ by analyzing
$\tilde{\theta}_{n}^{\mathtt{AD-BC,}\ast}$ directly, we find it more
insightful to obtain the consistency result by means of an argument which
highlights and exploits the relationship between $\tilde{\theta}%
_{n}^{\mathtt{AD-BC,}\ast}$ and $\hat{\theta}_{n}^{\mathtt{AD,}\ast}.$
Heuristically, $\tilde{\theta}_{n}^{\mathtt{AD-BC,}\ast}$ \textquotedblleft
should\textquotedblright\ satisfy $\left(
\ref{Bootstrap consistency: Alternative estimator}\right)  $ under Conditions
D, K, and B because the percentile interval associated with $\tilde{\theta
}_{n}^{\mathtt{AD-BC,}\ast}$ is identical to the percentile interval
associated with $\hat{\theta}_{n}^{\mathtt{AD,}\ast}.$\footnote{In generic
notation, the percentile interval associated with an estimator $\tilde{\theta
}_{n}^{\ast}$ is given by%
\[
\widetilde{\mathsf{CI}}_{n,1-\alpha}=\left[  \hat{\theta}_{n}-\tilde
{q}_{n,1-\alpha/2}^{\ast}~,~\hat{\theta}_{n}-\tilde{q}_{n,\alpha/2}^{\ast
}\right]  ,\qquad\tilde{q}_{n,a}^{\ast}=\inf\{q\in\mathbb{R}:\mathbb{P}%
_{n}^{\ast}[(\tilde{\theta}_{n}^{\ast}-\hat{\theta}_{n})\leq q]\geq a\}.
\]
} These heuristics can be made rigorous with the help of the equality%
\begin{align*}
& \sup_{t\in\mathbb{R}}\left\vert \mathbb{P}[\sqrt{n}(\hat{\theta}%
_{n}^{\mathtt{AD-BC}}-\theta_{0})\leq t]-\mathbb{P}_{n}^{\ast}[\sqrt{n}%
(\tilde{\theta}_{n}^{\mathtt{AD-BC,}\ast}-\hat{\theta}_{n}^{\mathtt{AD-BC}%
})\leq t]\right\vert \\
& =\sup_{t\in\mathbb{R}}\left\vert \mathbb{P}[\sqrt{n}(\hat{\theta}%
_{n}^{\mathtt{AD}}-\theta_{0})\leq t]-\mathbb{P}_{n}^{\ast}[\sqrt{n}%
(\hat{\theta}_{n}^{\mathtt{AD,}\ast}-\hat{\theta}_{n}^{\mathtt{AD}})\leq
t]\right\vert ,
\end{align*}
which implies in particular that $\tilde{\theta}_{n}^{\mathtt{AD-BC,}\ast}$
satisfies $\left(  \ref{Bootstrap consistency: Alternative estimator}\right)
$ if and only if $\hat{\theta}_{n}^{\mathtt{AD,}\ast}$ satisfies $\left(
\ref{Bootstrap consistency}\right)  .$ As a consequence, the fact
$\tilde{\theta}_{n}^{\mathtt{AD-BC,}\ast}$ satisfies $\left(
\ref{Bootstrap consistency: Alternative estimator}\right)  $ under Conditions
D, K, and B is simply a restatement of the bootstrap consistency result for
$\hat{\theta}_{n}^{\mathtt{AD,}\ast}.$

Turning next to $\hat{\theta}_{n}^{\mathtt{AD-LO}},$ our preferred
modification of this estimator is motivated by the observation that%
\[
\mathbb{P}[\tilde{f}_{i,n}^{\mathtt{LO}}(X_{i})=\hat{f}_{i,n}^{\mathtt{LO}%
}(X_{i})]=1,
\]
where%
\[
\tilde{f}_{i,n}^{\mathtt{LO}}(x)=\sum_{1\leq j\leq n}w_{ij,n}\tilde{K}%
_{n}(x-X_{j}),\qquad\tilde{K}_{n}(x)=%
\I
(x\neq0)K_{n}(x).
\]
An immediate implication of this observation is that%
\[
\mathbb{P}[\tilde{\theta}_{n}^{\mathtt{AD-LO}}=\hat{\theta}_{n}%
^{\mathtt{AD-LO}}]=1,\qquad\tilde{\theta}_{n}^{\mathtt{AD-LO}}=\frac{1}{n}%
\sum_{1\leq i\leq n}\tilde{f}_{i,n}^{\mathtt{LO}}(X_{i}).
\]
Nevertheless, unlike $\hat{\theta}_{n}^{\mathtt{AD-LO}}$ itself, the
modification $\tilde{\theta}_{n}^{\mathtt{AD-LO}}$ has a natural bootstrap
analog%
\[
\tilde{\theta}_{n}^{\mathtt{AD-LO,}\ast}=\frac{1}{n}\sum_{1\leq i\leq n}%
\tilde{f}_{i,n}^{\mathtt{LO,}\ast}(X_{i,n}^{\ast}),\qquad\tilde{f}%
_{i,n}^{\mathtt{LO,}\ast}(x)=\sum_{1\leq j\leq n}w_{ij,n}\tilde{K}%
_{n}(x-X_{j,n}^{\ast}),
\]
whose bias is small: Under Conditions D, K, and B,%
\[
\mathbb{E}_{n}^{\ast}[\tilde{\theta}_{n}^{\mathtt{AD-LO,}\ast}]=\frac{1}%
{n}\sum_{1\leq i\leq n}\tilde{f}_{n}^{\mathtt{LO}}(X_{i})=\tilde{\theta}%
_{n}^{\mathtt{AD-LO}}+o_{\mathbb{P}}(n^{-1/2}),\qquad\tilde{f}_{n}%
^{\mathtt{LO}}(x)=\frac{1}{n}\sum_{1\leq j\leq n}\tilde{K}_{n}(x-X_{j}).
\]
In fact, it can be shown that $\left(
\ref{Bootstrap consistency: Alternative estimator}\right)  $ is satisfied by
$\tilde{\theta}_{n}^{\mathtt{AD-LO,}\ast}$ under Conditions D, K, and B.

For cross-fit estimators, an arguably more attractive option is to construct a
bootstrap-based distributional approximation which employs a bootstrap
probability measure that is itself of cross-fit (i.e., split sample) type. To
illustrate the idea, we consider the simplest special case. When $B_{n}=2, $
the estimator $\hat{\theta}_{n}^{\mathtt{AD-LO}}$ reduces to%
\[
\hat{\theta}_{n}^{\mathtt{AD-CF}}=\frac{1}{n}\sum_{1\leq i\leq n}\hat{f}%
_{i,n}^{\mathtt{CF}}(X_{i}),
\]
where%
\[
\hat{f}_{i,n}^{\mathtt{CF}}(x)=\left\{
\begin{array}
[c]{cc}%
\frac{1}{n-\left\lfloor n/2\right\rfloor }\sum_{\left\lfloor n/2\right\rfloor
+1\leq j\leq n}K_{n}(x-X_{j}),\qquad & i\in\{1,\ldots,\left\lfloor
n/2\right\rfloor \}\\
\text{ } & \\
\frac{1}{\left\lfloor n/2\right\rfloor }\sum_{1\leq j\leq\left\lfloor
n/2\right\rfloor }K_{n}(x-X_{j}),\qquad & i\in\{\left\lfloor n/2\right\rfloor
+1,\ldots,n\}
\end{array}
\right.  .
\]
The $B_{n}=2$ version of the \textquotedblleft cross-fit
bootstrap\textquotedblright\ is defined as follows. Conditional on
$X_{1},\ldots,X_{n},$ let $X_{1,n}^{\star},\ldots,X_{n,n}^{\star}$ be mutually
independent with $X_{1,n}^{\star},\ldots,X_{\left\lfloor n/2\right\rfloor
,n}^{\star}$ being a random sample from the empirical distribution of
$X_{1},\ldots,X_{\left\lfloor n/2\right\rfloor }$ and $X_{\left\lfloor
n/2\right\rfloor +1,n}^{\star},\ldots,X_{n,n}^{\star}$ being a random sample
from the empirical distribution of $X_{\left\lfloor n/2\right\rfloor
+1},\ldots,X_{n}.$ Then,%
\[
\hat{\theta}_{n}^{\mathtt{AD-CF,}\star}=\frac{1}{n}\sum_{1\leq i\leq n}\hat
{f}_{i,n}^{\mathtt{CF,}\star}(X_{i,n}^{\star})
\]
is the corresponding cross-fit bootstrap\ version of $\hat{\theta}%
_{n}^{\mathtt{AD-CF}},$ where%
\[
\hat{f}_{i,n}^{\mathtt{CF,}\star}(x)=\left\{
\begin{array}
[c]{cc}%
\frac{1}{n-\left\lfloor n/2\right\rfloor }\sum_{\left\lfloor n/2\right\rfloor
+1\leq j\leq n}K_{n}(x-X_{j,n}^{\star}),\qquad & i\in\{1,\ldots,\left\lfloor
n/2\right\rfloor \}\\
\text{ } & \\
\frac{1}{\left\lfloor n/2\right\rfloor }\sum_{1\leq j\leq\left\lfloor
n/2\right\rfloor }K_{n}(x-X_{j,n}^{\star}),\qquad & i\in\{\left\lfloor
n/2\right\rfloor +1,\ldots,n\}
\end{array}
\right.  .
\]
The bootstrap distribution of $\hat{\theta}_{n}^{\mathtt{AD-CF,}\star}$ is
correctly centered in the sense that $\mathbb{E}_{n}^{\star}[\hat{\theta}%
_{n}^{\mathtt{AD-CF,}\star}]=\hat{\theta}_{n}^{\mathtt{AD-CF}},$ where
$\mathbb{E}_{n}^{\star}[\cdot]$ denotes the expected value computed under the
cross-fit bootstrap distribution. In fact, the bootstrap distribution
satisfies $\left(  \ref{Bootstrap consistency: Alternative measure}\right)  $
under Conditions D, K, and B.

As pointed out by a referee, yet another way of achieving consistency on the
part of a bootstrap-based distributional approximation is to center the
distribution of $\hat{\theta}_{n}^{\ast}$ at an estimator $\tilde{\theta}_{n}$
satisfying%
\begin{equation}
\sup_{t\in\mathbb{R}}\left\vert \mathbb{P}[\sqrt{n}(\hat{\theta}_{n}%
-\theta_{0})\leq t]-\mathbb{P}_{n}^{\ast}[\sqrt{n}(\hat{\theta}_{n}^{\ast
}-\tilde{\theta}_{n})\leq t]\right\vert =o_{\mathbb{P}}%
(1).\label{Bootstrap consistency: Alternative centering}%
\end{equation}
Because the estimators under consideration here all satisfiy $\left(
\ref{Efficiency: Noise condition}\right)  $ and $\left(
\ref{Bootstrap consistency: Noise condition}\right)  ,$ the following analog
of $\left(  \ref{Bootstrap consistency: Bias condition}\right)  $ is
sufficient for $\left(  \ref{Bootstrap consistency: Alternative centering}%
\right)  :$%
\[
\sqrt{n}(\mathbb{E}_{n}^{\ast}[\hat{\theta}_{n}^{\ast}]-\tilde{\theta}%
_{n})=\sqrt{n}(\mathbb{E}[\hat{\theta}_{n}]-\theta_{0})+o_{\mathbb{P}}(1).
\]
As already mentioned in connection with $\left(
\ref{Bootstrap consistency: Bias condition}\right)  ,$ the displayed condition
satisfied by $\tilde{\theta}_{n}=\hat{\theta}_{n}$ in the case of $\hat
{\theta}_{n}^{\mathtt{AD}}.$ For the other estimators (i.e., for $\hat{\theta
}_{n}^{\mathtt{AD-BC}},$ $\hat{\theta}_{n}^{\mathtt{AD-GJ}},$ and $\hat
{\theta}_{n}^{\mathtt{AD-LO}}$), because they satisfy the bias condition
$\left(  \ref{Efficiency: Bias condition}\right)  ,$ the displayed condition
is satisfied by $\tilde{\theta}_{n}=\mathbb{E}_{n}^{\ast}[\hat{\theta}%
_{n}^{\ast}].$

\section{Alternative Estimators\label{[Section] Alternative Estimators}}

This section considers two alternative classes of estimators. The first class
is motivated by the integrated squared density representation%
\[
\theta_{0}=\int_{\mathbb{R}^{d}}f_{0}\left(  x\right)  ^{2}dx,
\]
an interesting feature of which is that it involves a nonlinear functional of
$f_{0}.$ The second class is motivated by the representation%
\[
\theta_{0}=2\mathbb{E}[f_{0}(X)]-\int_{\mathbb{R}^{d}}f_{0}\left(  x\right)
^{2}dx,
\]
an interesting feature of which is that it is \textquotedblleft locally
robust\textquotedblright/\textquotedblleft Neyman orthogonal\textquotedblright%
\ (in the terminology of
\cite{Chernozhukov-Escanciano-Ichimura-Newey-Robins_2020_LocallyRobust}).

\subsection{Integrated Squared Density
Estimators\label{[Subsection] Integrated Squared Density Estimators}}

A kernel-based plug-in integrated squared density estimator is%
\[
\hat{\theta}_{n}^{\mathtt{ISD}}=\int_{\mathbb{R}^{d}}\hat{f}_{n}\left(
x\right)  ^{2}dx.
\]
Like $\hat{\theta}_{n}^{\mathtt{AD}},$ this estimator has a (potentially)
nonnegligible bias: Under Conditions D, K, and B,%
\[
\mathbb{E}[\hat{\theta}_{n}^{\mathtt{ISD}}]-\theta_{0}=\frac{\int%
_{\mathbb{R}^{d}}K(u)^{2}du}{nh_{n}^{d}}+o(n^{-1/2}),
\]
where the first term is a \textquotedblleft nonlinearity\textquotedblright%
\ bias term (in the terminology of \cite{Cattaneo-Crump-Jansson_2013_JASA})
attributable to the fact that $\hat{\theta}_{n}^{\mathtt{ISD}}$ is a nonlinear
functional of $\hat{f}_{n}.$

The nonlinearity bias of $\hat{\theta}_{n}^{\mathtt{ISD}}$ is easily
avoidable, a simple bias-corrected version of $\hat{\theta}_{n}^{\mathtt{ISD}%
}$ being%
\[
\hat{\theta}_{n}^{\mathtt{ISD-BC}}=\hat{\theta}_{n}^{\mathtt{ISD}}-\frac
{\int_{\mathbb{R}^{d}}K(u)^{2}du}{nh_{n}^{d}}.
\]
Similarly, because the nonlinearity bias of $\hat{\theta}_{n}^{\mathtt{ISD}}$
is proportional to $1/(nh_{n}^{d}),$ the following generalized jackknife
version of $\hat{\theta}_{n}^{\mathtt{ISD}}$ is an efficient estimator of
$\theta_{0}:$%
\[
\hat{\theta}_{n}^{\mathtt{ISD-GJ}}=\frac{1}{1-c^{d}}\hat{\theta}%
_{n}^{\mathtt{ISD}}-\frac{c^{d}}{1-c^{d}}\hat{\theta}_{n}^{\mathtt{ISD}%
}(ch_{n}),
\]
where $c\neq1$ is a user-chosen constant and where $\hat{\theta}%
_{n}^{\mathtt{ISD}}(ch_{n})$ denotes the version of $\hat{\theta}%
_{n}^{\mathtt{ISD}}$ associated with the bandwidth $ch_{n}.$

On the other hand, because the source of the nonlinearity bias of $\hat
{\theta}_{n}^{\mathtt{ISD}}$ is different from the source of the leave in bias
of $\hat{\theta}_{n}^{\mathtt{AD}},$ there is no particular reason to expect
leave out estimators of the form%
\[
\hat{\theta}_{n}^{\mathtt{ISD-LO}}=\frac{1}{n}\sum_{1\leq i\leq n}%
\int_{\mathbb{R}^{d}}\hat{f}_{i,n}^{\mathtt{LO}}(x)^{2}dx
\]
to have favorable bias properties. Indeed, under Conditions D, K, and B and
assuming $B_{n}$ is proportional to $n,$ we have:\footnote{More generally
(i.e., whether or not $B_{n}$ is proportional to $n$), it is shown in the
proof of Theorem \ref{[Theorem] Efficiency (ISD)} that the bias expansion is
of the form%
\[
\mathbb{E}[\hat{\theta}_{n}^{\mathtt{ISD-LO}}]-\theta_{0}=\eta_{n}\frac
{\int_{\mathbb{R}^{d}}K(u)^{2}du}{nh_{n}^{d}}+o(n^{-1/2}),
\]
where $\eta_{n}\geq1$ is bounded.}%
\begin{equation}
\mathbb{E}[\hat{\theta}_{n}^{\mathtt{ISD-LO}}]-\theta_{0}=\frac{1}%
{1-B_{n}^{-1}}\frac{\int_{\mathbb{R}^{d}}K(u)^{2}du}{nh_{n}^{d}}%
+o(n^{-1/2}),\label{Leave-Out ISD estimator: Bias}%
\end{equation}
so the nonlinearity bias of $\hat{\theta}_{n}^{\mathtt{ISD-LO}}$ is
nonnegligible (and no smaller than that of $\hat{\theta}_{n}^{\mathtt{ISD}}$).

Nevertheless, because $\theta_{0}$ is a quadratic functional of $f_{0},$ the
method of \textquotedblleft doubly cross-fitting\textquotedblright\ (in the
terminology of \cite{Newey-Robins_2018_CrossFitting}) can be used to construct
an estimator which is free of nonlinearity bias and can be implemented without
knowledge of the functional form of the nonlinearity bias. One such estimator
is%
\[
\hat{\theta}_{n}^{\mathtt{ISD-DCF}}=\int_{\mathbb{R}^{d}}\hat{f}%
_{1,n}^{\mathtt{CF}}\left(  x\right)  \hat{f}_{n,n}^{\mathtt{CF}}\left(
x\right)  dx,
\]
whose bias turns out to be negligible under Conditions D, K, and B.

Under Conditions D, K, and B$^{-}$, the estimators $\hat{\theta}%
_{n}^{\mathtt{ISD}},$ $\hat{\theta}_{n}^{\mathtt{ISD-BC}},$ $\hat{\theta}%
_{n}^{\mathtt{ISD-GJ}},$ $\hat{\theta}_{n}^{\mathtt{ISD-LO}},$ and
$\hat{\theta}_{n}^{\mathtt{ISD-DCF}}$ all satisfy $\left(
\ref{Efficiency: Noise condition}\right)  .$ As a consequence, we obtain the
following integrated squared density counterpart of Theorem
\ref{[Theorem] Efficiency}.

\begin{theorem}
\label{[Theorem] Efficiency (ISD)}Suppose Conditions D, K, and B are
satisfied. Then $\hat{\theta}_{n}^{\mathtt{ISD-BC}},$ $\hat{\theta}%
_{n}^{\mathtt{ISD-GJ}},$ and $\hat{\theta}_{n}^{\mathtt{ISD-DCF}}$ satisfy
$\left(  \ref{Efficiency}\right)  .$ If Condition B is strengthened to
Condition B$^{+} $, then $\hat{\theta}_{n}^{\mathtt{ISD}}$ and $\hat{\theta
}_{n}^{\mathtt{ISD-LO}}$ satisfy $\left(  \ref{Efficiency}\right)  .$
\end{theorem}

An integrated squared density counterpart of Theorem
\ref{[Theorem] Bootstrap consistency} is also available. Under Conditions D,
K, and B, if $\hat{\theta}_{n}\in\{\hat{\theta}_{n}^{\mathtt{ISD}},\hat
{\theta}_{n}^{\mathtt{ISD-BC}},\hat{\theta}_{n}^{\mathtt{ISD-LO}},\hat{\theta
}_{n}^{\mathtt{ISD-DCF}}\},$ then its bootstrap analog satisfies $\left(
\ref{Bootstrap consistency: Noise condition}\right)  $ and has a bias of the
form%
\[
\mathbb{E}_{n}^{\ast}[\hat{\theta}_{n}^{\ast}]-\hat{\theta}_{n}=\frac
{\int_{\mathbb{R}^{d}}K(u)^{2}du}{nh_{n}^{d}}+o_{\mathbb{P}}(n^{-1/2}),
\]
so $\left(  \ref{Bootstrap consistency}\right)  $ is satisfied if (and only
if)%
\[
\mathbb{E}[\hat{\theta}_{n}]-\theta_{0}=\frac{\int_{\mathbb{R}^{d}}K(u)^{2}%
du}{nh_{n}^{d}}+o(n^{-1/2}).
\]
\newline The latter condition is satisfied by $\hat{\theta}_{n}^{\mathtt{ISD}%
},$ but violated by $\hat{\theta}_{n}^{\mathtt{ISD-BC}}$ and $\hat{\theta}%
_{n}^{\mathtt{ISD-DCF}}.$ In the case of $\hat{\theta}_{n}^{\mathtt{ISD-LO}},$
it follows from $\left(  \ref{Leave-Out ISD estimator: Bias}\right)  $ that
the condition is satisfied when $B_{n}=n$ (i.e., when $\hat{\theta}%
_{n}^{\mathtt{ISD-LO}}$ is a leave-one-out estimator), but violated when
$B_{n}$ is fixed (i.e., when $\hat{\theta}_{n}^{\mathtt{ISD-LO}}$ is a
cross-fit estimator).

\begin{theorem}
\label{[Theorem] Bootstrap consistency (ISD)}Suppose Conditions D, K, and B
are satisfied. Then $\hat{\theta}_{n}^{\mathtt{ISD,}\ast}$ and $\hat{\theta
}_{n}^{\mathtt{ISD-GJ,}\ast}$ satisfy $\left(  \ref{Bootstrap consistency}%
\right)  .$ If $B_{n}=n,$ then $\hat{\theta}_{n}^{\mathtt{ISD-LO,}\ast}$
satisfies $\left(  \ref{Bootstrap consistency}\right)  .$ If Condition B is
strengthened to Condition B$^{+}$, then $\hat{\theta}_{n}^{\mathtt{ISD-BC,}%
\ast},$ $\hat{\theta}_{n}^{\mathtt{ISD-LO,}\ast},$ and $\hat{\theta}%
_{n}^{\mathtt{ISD-DCF,}\ast}$ satisfy $\left(  \ref{Bootstrap consistency}%
\right)  . $
\end{theorem}

In important respects, the results reported in Theorems
\ref{[Theorem] Efficiency (ISD)} and
\ref{[Theorem] Bootstrap consistency (ISD)} are in qualitative agreement with
those reported in Theorems \ref{[Theorem] Efficiency} and
\ref{[Theorem] Bootstrap consistency}. In particular, we find that in spite of
being inefficient the simple plug-in estimator achieves bootstrap consistency
under conditions that are weaker than those required for efficient estimators
to achieve bootstrap consistency. The most notable difference between the
integrated squared density and average derivative estimators is probably that
in the case of integrated squared density estimators, the cross-fit estimator
is demonstrably worse than the plug-in estimator, satisfying neither $\left(
\ref{Efficiency}\right)  $ nor $\left(  \ref{Bootstrap consistency}\right)  .$

As was the case with the average density estimators, the integrated squared
density estimators can all be shown to satisfy the bootstrap variance
consistency property $(\ref{Bootstrap variance consistency})$ under Conditions
D, K, and B. The estimators $\hat{\theta}_{n}^{\mathtt{ISD-BC}},$ $\hat
{\theta}_{n}^{\mathtt{ISD-GJ}},$ and $\hat{\theta}_{n}^{\mathtt{ISD-DCF}}$
therefore enjoy the property that the intervals $\mathsf{CI}_{n,1-\alpha
}^{\mathtt{N}}$ based on the bootstrap variance estimator are consistent
(indeed, efficient) under Conditions D, K, and B.

For completeness, we conclude this subsection by briefly discussing integrated
squared density versions of $\left(
\ref{Bootstrap consistency: Alternative estimator}\right)  ,$ $\left(
\ref{Bootstrap consistency: Alternative measure}\right)  $, and $\left(
\ref{Bootstrap consistency: Alternative centering}\right)  .$ In what follows,
suppose Conditions D, K, and B are satisfied. A bias-corrected version of
$\hat{\theta}_{n}^{\mathtt{ISD-BC,}\ast}$ is given by%
\[
\tilde{\theta}_{n}^{\mathtt{ISD-BC,}\ast}=\hat{\theta}_{n}^{\mathtt{ISD-BC,}%
\ast}-\frac{\int_{\mathbb{R}^{d}}K(u)^{2}du}{nh_{n}^{d}}.
\]
In perfect analogy with $\tilde{\theta}_{n}^{\mathtt{AD-BC,}\ast},$ this
estimator satisfies $\left(
\ref{Bootstrap consistency: Alternative estimator}\right)  $ and the
associated percentile interval is identical to the percentile interval
associated with $\hat{\theta}_{n}^{\mathtt{ISD,}\ast}.$ Next,%
\[
\tilde{\theta}_{n}^{\mathtt{ISD-LO,}\ast}=\frac{1}{n}\sum_{1\leq i\leq n}%
\int_{\mathbb{R}^{d}}\tilde{f}_{i,n}^{\mathtt{LO,}\ast}(x)^{2}dx
\]
is an integrated squared density counterpart of $\tilde{\theta}_{n}%
^{\mathtt{AD-LO,}\ast}.$ Because $\tilde{\theta}_{n}^{\mathtt{ISD-LO,}\ast
}=\hat{\theta}_{n}^{\mathtt{ISD-LO,}\ast},$ this estimator satisfies $\left(
\ref{Bootstrap consistency: Alternative estimator}\right)  $ when $B_{n}=n,$
but not when $B_{n}$ is fixed. On the other hand, the cross-fit bootstrap can
be used when $B_{n}$ is fixed. As before, suppose $B_{n}=2$ for specificity.
In that case, $\hat{\theta}_{n}^{\mathtt{ISD-LO}}$ reduces to%
\[
\hat{\theta}_{n}^{\mathtt{ISD-CF}}=\frac{1}{n}\sum_{1\leq i\leq n}%
\int_{\mathbb{R}^{d}}\hat{f}_{i,n}^{\mathtt{CF}}(x)^{2}dx
\]
and it can be shown that%
\[
\hat{\theta}_{n}^{\mathtt{ISD-CF,}\star}=\frac{1}{n}\sum_{1\leq i\leq n}%
\int_{\mathbb{R}^{d}}\hat{f}_{i,n}^{\mathtt{CF,}\star}(x)^{2}dx
\]
satisfies $\left(  \ref{Bootstrap consistency: Alternative measure}\right)  .$
Similarly, the distribution of $\hat{\theta}_{n}^{\mathtt{ISD-DCF}}$ can be
approximated using%
\[
\hat{\theta}_{n}^{\mathtt{ISD-DCF,}\star}=\int_{\mathbb{R}^{d}}\hat{f}%
_{1,n}^{\mathtt{CF,}\star}\left(  x\right)  \hat{f}_{n,n}^{\mathtt{CF,}\star
}\left(  x\right)  dx,
\]
as that estimator satisfies $\left(
\ref{Bootstrap consistency: Alternative measure}\right)  .$ Finally, the
property $\left(  \ref{Bootstrap consistency: Alternative centering}\right)  $
holds for $\hat{\theta}_{n}^{\mathtt{ISD-BC}},$ $\hat{\theta}_{n}%
^{\mathtt{ISD-GJ}},$ and $\hat{\theta}_{n}^{\mathtt{ISD-DCF}}$ when
$\tilde{\theta}_{n}=\mathbb{E}_{n}^{\ast}[\hat{\theta}_{n}^{\ast}]$ (and for
$\hat{\theta}_{n}^{\mathtt{ISD}},$ $\hat{\theta}_{n}^{\mathtt{ISD-GJ}},$ and
the leave-one-out version of $\hat{\theta}_{n}^{\mathtt{ISD-LO}}$ when
$\tilde{\theta}_{n}=\hat{\theta}_{n}$).

\subsection{Locally Robust
Estimators\label{[Subsection] Locally Robust Estimators}}

A locally robust kernel-based plug-in\ estimator of $\theta_{0}$ is%
\[
\hat{\theta}_{n}^{\mathtt{LR}}=\frac{2}{n}\sum_{1\leq i\leq n}\hat{f}%
_{n}(X_{i})-\int_{\mathbb{R}^{d}}\hat{f}_{n}\left(  x\right)  ^{2}%
dx=2\hat{\theta}_{n}^{\mathtt{AD}}-\hat{\theta}_{n}^{\mathtt{ISD}}.
\]
Because $\hat{\theta}_{n}^{\mathtt{LR}}$ is a linear combination of
$\hat{\theta}_{n}^{\mathtt{AD}}$ and $\hat{\theta}_{n}^{\mathtt{ISD}},$ its
properties follow directly from the results obtained in the previous sections,
as do the properties of estimators such as%
\[
\hat{\theta}_{n}^{\mathtt{LR-BC}}=2\hat{\theta}_{n}^{\mathtt{AD-BC}}%
-\hat{\theta}_{n}^{\mathtt{ISD-BC}},
\]%
\[
\hat{\theta}_{n}^{\mathtt{LR-GJ}}=2\hat{\theta}_{n}^{\mathtt{AD-GJ}}%
-\hat{\theta}_{n}^{\mathtt{ISD-GJ}},
\]
and%
\[
\hat{\theta}_{n}^{\mathtt{LR-LO}}=2\hat{\theta}_{n}^{\mathtt{AD-LO}}%
-\hat{\theta}_{n}^{\mathtt{ISD-LO}},
\]
the cross-fit version of the latter being the only estimator (in this paper)
satisfying both of the defining properties of the \textquotedblleft
double/debiased machine learning\textquotedblright\ estimators proposed by
\cite{Chernozhukov-Chetverikov-Demirer-Duflo-Hansen-Newey-Robins_2018_EctJ}.

Once again, the results are in qualitative agreement with those reported in
Theorems \ref{[Theorem] Efficiency} and \ref{[Theorem] Bootstrap consistency}.

\begin{theorem}
\label{[Theorem] Efficiency (LR)}Suppose Conditions D, K, and B are satisfied.
Then $\hat{\theta}_{n}^{\mathtt{LR-BC}}$ and $\hat{\theta}_{n}^{\mathtt{LR-GJ}%
}$ satisfy $\left(  \ref{Efficiency}\right)  .$ If Condition B is strengthened
to Condition B$^{+}$, then $\hat{\theta}_{n}^{\mathtt{LR}}$ and $\hat{\theta
}_{n}^{\mathtt{LR-LO}}$ satisfy $\left(  \ref{Efficiency}\right)  .$
\end{theorem}

\begin{theorem}
\label{[Theorem] Bootstrap consistency (LR)}Suppose Conditions D, K, and B are
satisfied. Then $\hat{\theta}_{n}^{\mathtt{LR,}\ast}$ and $\hat{\theta}%
_{n}^{\mathtt{LR-GJ,}\ast}$ satisfy $\left(  \ref{Bootstrap consistency}%
\right)  .$ If Condition B is strengthened to Condition B$^{+}$, then
$\hat{\theta}_{n}^{\mathtt{LR-BC,}\ast}$ and $\hat{\theta}_{n}%
^{\mathtt{LR-LO,}\ast}$ satisfy $\left(  \ref{Bootstrap consistency}\right)
.$
\end{theorem}

Rather than spelling out those locally robust versions of $\left(
\ref{Bootstrap consistency: Alternative estimator}\right)  ,$ $\left(
\ref{Bootstrap consistency: Alternative measure}\right)  $, and $\left(
\ref{Bootstrap consistency: Alternative centering}\right)  $ that follow
directly from our earlier results, it seems more constructive to mention a
feature of local robustness that is particularly useful for boostrap purposes.
As pointed out by \cite{Belloni-Chernozhukov-FernandezVal-Hansen_2017_ECMA}, a
notable feature of locally robust moment conditions is that in two-step
estimation settings one does not need to recompute the first step estimator in
each iteration of the bootstrap. In the case of $\hat{\theta}_{n}%
^{\mathtt{LR}},$ this implies that $\left(
\ref{Bootstrap consistency: Alternative estimator}\right)  $ can be achieved
with the help of%
\[
\tilde{\theta}_{n}^{\mathtt{LR,}\ast}=\frac{2}{n}\sum_{1\leq i\leq n}\hat
{f}_{n}(X_{i,n}^{\ast})-\int_{\mathbb{R}^{d}}\hat{f}_{n}\left(  x\right)
^{2}dx,
\]
a computationally attractive feature of which is that $\hat{f}_{n}$ is kept
fixed across bootstrap repetitions. Perhaps more importantly (for our purposes
at least), the fact that $\hat{f}_{n}$ is kept fixed actually makes it easier
to achieve $\left(  \ref{Bootstrap consistency: Alternative estimator}\right)
$ also in the case of debiased estimators. For instance,%
\[
\tilde{\theta}_{n}^{\mathtt{LR-BC,}\ast}=\frac{2}{n}\sum_{1\leq i\leq n}%
\hat{f}_{n}(X_{i,n}^{\ast})-\int_{\mathbb{R}^{d}}\hat{f}_{n}\left(  x\right)
^{2}dx-\frac{2K(0)-\int_{\mathbb{R}^{d}}K(u)^{2}du}{nh_{n}^{d}}%
\]
satisfies $\left(  \ref{Bootstrap consistency: Alternative estimator}\right)
$ under Conditions D, K, and B.

\section{Proofs\label{[Section] Proofs}}

\subsection{Hoeffding Decompositions}

Each of the estimators studied in this paper has a $V$-statistic-type
representation of the form%
\[
\hat{\theta}_{n}=\frac{1}{n^{2}}\sum_{1\leq i,j\leq n}V_{ij,n},
\]
where $V_{ij,n}$ depends on $X_{1},\ldots,X_{n}$ only through $(X_{i},X_{j}).$
The proofs of Theorems \ref{[Theorem] Efficiency},
\ref{[Theorem] Efficiency (ISD)}, and \ref{[Theorem] Efficiency (LR)} are
based on the associated Hoeffding decomposition of $\hat{\theta}_{n}%
-\theta_{0}$ given by%
\begin{equation}
\hat{\theta}_{n}-\theta_{0}=\beta_{n}+\frac{1}{n}\sum_{1\leq i\leq n}%
L_{i,n}+\frac{2}{n(n-1)}\sum_{1\leq i,j\leq n,i<j}W_{ij,n}%
,\label{Hoeffding decomposition}%
\end{equation}
where, defining $\bar{V}_{ij,n}=(V_{ij,n}+V_{ji,n})/2,$%
\begin{align*}
\beta_{n}  & =\mathbb{E}[\hat{\theta}_{n}]-\theta_{0}\\
& =\frac{1}{n}\left\{  \frac{1}{n}\sum_{1\leq i\leq n}\mathbb{E}%
[V_{ii,n}]\right\}  +\left(  1-\frac{1}{n}\right)  \left\{  \frac{2}%
{n(n-1)}\sum_{1\leq i,j\leq n,i<j}\mathbb{E}[\bar{V}_{ij,n}]\right\}
-\theta_{0},
\end{align*}%
\begin{align*}
L_{i,n}  & =n\{\mathbb{E}[\hat{\theta}_{n}|X_{i}]-\mathbb{E}[\hat{\theta}%
_{n}]\}\\
& =\frac{1}{n}\{V_{ii,n}-\mathbb{E}[V_{ii,n}]\}+\frac{1}{n-1}\sum_{1\leq j\leq
n,j\neq i}2\frac{n-1}{n}\{\mathbb{E}[\bar{V}_{ij,n}|X_{i}]-\mathbb{E}[\bar
{V}_{ij,n}]\},
\end{align*}%
\begin{align*}
W_{ij,n}  & =\frac{n(n-1)}{2}\{\mathbb{E}[\hat{\theta}_{n}|X_{i}%
,X_{j}]-\mathbb{E}[\hat{\theta}_{n}|X_{i}]-\mathbb{E}[\hat{\theta}_{n}%
|X_{j}]+\mathbb{E}[\hat{\theta}_{n}]\}\\
& =\frac{n-1}{n}\{\bar{V}_{ij,n}-\mathbb{E}[\bar{V}_{ij,n}|X_{i}%
]-\mathbb{E}[\bar{V}_{ij,n}|X_{j}]+\mathbb{E}[\bar{V}_{ij,n}]\}.
\end{align*}

By construction, $L_{i,n}$ and $W_{ij,n}$ depend on $X_{1},\ldots,X_{n}$ only
through $X_{i}$ and $(X_{i},X_{j}),$ respectively, and satisfy, for each
$1\leq i,j\leq n$ with $i\neq j,$%
\[
\mathbb{E}[L_{i,n}]=\mathbb{E}[W_{ij,n}|X_{i}]=\mathbb{E}[W_{ij,n}|X_{j}]=0.
\]
Moreover, if the $V_{ij,n}$ satisfy $V_{ii,n}=\delta_{n}$ and $\mathbb{E}%
[V_{ij,n}]=\theta_{n},$ then the bias is of the form%
\[
\beta_{n}=\frac{\delta_{n}}{n}+\theta_{n}-\theta_{0}-\frac{\theta_{n}}{n}.
\]
If also $V_{ij,n}=V_{ji,n}$ and $\mathbb{E}[V_{ij,n}|X_{i}]=f_{n}(X_{i}),$
then%
\[
L_{i,n}=2\frac{n-1}{n}\{f_{n}(X_{i})-\theta_{n}\},\qquad W_{ij,n}=\frac
{n-1}{n}\{V_{ij,n}-f_{n}(X_{i})-f_{n}(X_{j})+\theta_{n}\}.
\]

A bootstrap analog of $\left(  \ref{Hoeffding decomposition}\right)  $ will be
employed in the proofs of Theorems \ref{[Theorem] Bootstrap consistency},
\ref{[Theorem] Bootstrap consistency (ISD)}, and
\ref{[Theorem] Bootstrap consistency (LR)}. To state it, suppose%
\[
\hat{\theta}_{n}^{\ast}=\frac{1}{n^{2}}\sum_{1\leq i,j\leq n}V_{ij,n}^{\ast},
\]
where $V_{ij,n}^{\ast}$ depends on $X_{1,n}^{\ast},\ldots,X_{n,n}^{\ast}$ only
through $(X_{i,n}^{\ast},X_{j,n}^{\ast}).$ Then%
\begin{equation}
\hat{\theta}_{n}^{\ast}-\hat{\theta}_{n}=\beta_{n}^{\ast}+\frac{1}{n}%
\sum_{1\leq i\leq n}L_{i,n}^{\ast}+\frac{2}{n(n-1)}\sum_{1\leq i,j\leq
n,i<j}W_{ij,n}^{\ast},\label{Hoeffding decomposition (bootstrap)}%
\end{equation}
where, defining $\bar{V}_{ij,n}^{\ast}=(V_{ij,n}^{\ast}+V_{ji,n}^{\ast})/2,$%
\begin{align*}
\beta_{n}^{\ast}  & =\mathbb{E}_{n}^{\ast}[\hat{\theta}_{n}^{\ast}%
]-\hat{\theta}_{n}\\
& =\frac{1}{n}\left\{  \frac{1}{n}\sum_{1\leq i\leq n}\mathbb{E}_{n}^{\ast
}[V_{ii,n}^{\ast}]\right\}  +\left(  1-\frac{1}{n}\right)  \left\{  \frac
{2}{n(n-1)}\sum_{1\leq i,j\leq n,i<j}\mathbb{E}_{n}^{\ast}[\bar{V}%
_{ij,n}^{\ast}]\right\}  -\hat{\theta}_{n},
\end{align*}%
\begin{align*}
L_{i,n}^{\ast}  & =n\{\mathbb{E}_{n}^{\ast}[\hat{\theta}_{n}^{\ast}%
|X_{i,n}^{\ast}]-\mathbb{E}_{n}^{\ast}[\hat{\theta}_{n}^{\ast}]\}\\
& =\frac{1}{n}\{V_{ii,n}^{\ast}-\mathbb{E}_{n}^{\ast}[V_{ii,n}^{\ast}%
]\}+\frac{1}{n-1}\sum_{1\leq j\leq n,j\neq i}2\frac{n-1}{n}\{\mathbb{E}%
_{n}^{\ast}[\bar{V}_{ij,n}^{\ast}|X_{i,n}^{\ast}]-\mathbb{E}_{n}^{\ast}%
[\bar{V}_{ij,n}^{\ast}]\},
\end{align*}%
\begin{align*}
W_{ij,n}^{\ast}  & =\frac{n(n-1)}{2}\{\mathbb{E}_{n}^{\ast}[\hat{\theta}%
_{n}^{\ast}|X_{i,n}^{\ast},X_{j,n}^{\ast}]-\mathbb{E}_{n}^{\ast}[\hat{\theta
}_{n}^{\ast}|X_{i,n}^{\ast}]-\mathbb{E}_{n}^{\ast}[\hat{\theta}_{n}^{\ast
}|X_{j,n}^{\ast}]+\mathbb{E}_{n}^{\ast}[\hat{\theta}_{n}^{\ast}]\}\\
& =\frac{n-1}{n}\{\bar{V}_{ij,n}^{\ast}-\mathbb{E}_{n}^{\ast}[\bar{V}%
_{ij,n}^{\ast}|X_{i,n}^{\ast}]-\mathbb{E}_{n}^{\ast}[\bar{V}_{ij,n}^{\ast
}|X_{j,n}^{\ast}]+\mathbb{E}_{n}^{\ast}[\bar{V}_{ij,n}^{\ast}]\}.
\end{align*}

By construction, $L_{i,n}^{\ast}$ and $W_{ij,n}^{\ast}$ depend on
$X_{1,n}^{\ast},\ldots,X_{n,n}^{\ast}$ only through $X_{i,n}^{\ast}$ and
$(X_{i,n}^{\ast},X_{j,n}^{\ast}),$ respectively, and satisfy, for each $1\leq
i,j\leq n$ with $i\neq j,$%
\[
\mathbb{E}_{n}^{\ast}[L_{i,n}^{\ast}]=\mathbb{E}_{n}^{\ast}[W_{ij,n}^{\ast
}|X_{i,n}^{\ast}]=\mathbb{E}_{n}^{\ast}[W_{ij,n}^{\ast}|X_{j,n}^{\ast}]=0.
\]
Moreover, if the $V_{ij,n}^{\ast}$ satisfy $V_{ii,n}^{\ast}=\delta_{n}^{\ast}$
and $\mathbb{E}_{n}^{\ast}[V_{ij,n}^{\ast}]=\theta_{n}^{\ast},$ then the
bootstrap bias is of the form%
\[
\beta_{n}^{\ast}=\frac{\delta_{n}^{\ast}}{n}+\theta_{n}^{\ast}-\hat{\theta
}_{n}-\frac{\theta_{n}^{\ast}}{n}.
\]
If also $V_{ij,n}^{\ast}=V_{ji,n}^{\ast}$ and $\mathbb{E}_{n}^{\ast}%
[V_{ij,n}^{\ast}|X_{i,n}^{\ast}]=f_{n}^{\ast}(X_{i,n}^{\ast}),$ then%
\[
L_{i,n}^{\ast}=2\frac{n-1}{n}\{f_{n}^{\ast}(X_{i,n}^{\ast})-\theta_{n}^{\ast
}\},\qquad W_{ij,n}^{\ast}=\frac{n-1}{n}\{V_{ij,n}^{\ast}-f_{n}^{\ast}%
(X_{i,n}^{\ast})-f_{n}^{\ast}(X_{j,n}^{\ast})+\theta_{n}^{\ast}\}.
\]

\subsection{Proof of Theorem \ref{[Theorem] Efficiency}}

The estimators $\hat{\theta}_{n}^{\mathtt{AD}}$ and $\hat{\theta}%
_{n}^{\mathtt{AD-LO}}$ both have Hoeffding decompositions of the form $\left(
\ref{Hoeffding decomposition}\right)  ,$ with%
\[
L_{i,n}=\lambda_{i,n}L_{n}^{\mathtt{AD}}(X_{i})\text{\qquad and\qquad}%
W_{ij,n}=\omega_{ij,n}W_{n}^{\mathtt{AD}}(X_{i},X_{j}),
\]
where $\lambda_{i,n}$ and $\omega_{ij,n}$ are (non-random) estimator-specific
weights, while%
\[
L_{n}^{\mathtt{AD}}(x)=2\{f_{n}^{\mathtt{AD}}(x)-\theta_{n}^{\mathtt{AD}}\},
\]%
\[
W_{n}^{\mathtt{AD}}(x_{1},x_{2})=K_{n}(x_{1}-x_{2})-f_{n}^{\mathtt{AD}}%
(x_{1})-f_{n}^{\mathtt{AD}}(x_{2})+\theta_{n}^{\mathtt{AD}},
\]
where%

\[
f_{n}^{\mathtt{AD}}\left(  x\right)  =\mathbb{E}[K_{n}(x-X)]=\int%
_{\mathbb{R}^{d}}K(u)f_{0}(x+uh_{n})du,
\]%
\[
\theta_{n}^{\mathtt{AD}}=\mathbb{E}[f_{n}^{\mathtt{AD}}(X)]=\int%
_{\mathbb{R}^{d}}f_{n}^{\mathtt{AD}}(x)f_{0}(x)dx.
\]

To be specific, in the case of%
\[
\hat{\theta}_{n}^{\mathtt{AD}}=\frac{1}{n}\sum_{1\leq i\leq n}\hat{f}%
_{n}(X_{i})=\frac{1}{n^{2}}\sum_{1\leq i,j\leq n}K_{n}(X_{i}-X_{j}),
\]
each $\lambda_{i,n}$ and $\omega_{ij,n}$ is given by $1-n^{-1},$ while the
weights for%
\[
\hat{\theta}_{n}^{\mathtt{AD-LO}}=\frac{1}{n}\sum_{1\leq i\leq n}\hat{f}%
_{i,n}^{\mathtt{LO}}(X_{i})=\frac{1}{n^{2}}\sum_{1\leq i,j\leq n}%
nw_{ij,n}K_{n}(X_{i}-X_{j})
\]
are of the form%
\[
\lambda_{i,n}=\sum_{1\leq j\leq n}\bar{w}_{ij,n},\qquad\omega_{ij,n}%
=(n-1)\bar{w}_{ij,n},\qquad\bar{w}_{ij,n}=(w_{ij,n}+w_{ji,n})/2.
\]

In both cases, the weights satisfy%
\begin{equation}
\max_{1\leq i\leq n}(\lambda_{i,n}-1)^{2}=o(1)\label{Linear term: weights}%
\end{equation}
and%
\begin{equation}
\max_{1\leq i<j\leq n}\omega_{ij,n}^{2}=O(1).\label{Quadratic term: weights}%
\end{equation}
It therefore follows from simple moment calculations that the estimators
satisfy $\left(  \ref{Efficiency: Noise condition}\right)  $ if%
\begin{equation}
\frac{1}{n}\mathbb{E}[W_{n}^{\mathtt{AD}}(X_{1},X_{2})^{2}]\rightarrow
0\label{Quadratic term: variance}%
\end{equation}
and if%
\begin{equation}
\mathbb{E}[\{L_{n}^{\mathtt{AD}}(X)-L_{0}(X)\}^{2}]\rightarrow
0.\label{Linear term: mean square convergence}%
\end{equation}

Suppose Conditions D and K are satisfied. Then $\left(
\ref{Quadratic term: variance}\right)  $ holds if $nh_{n}^{d}\rightarrow
\infty,$ because then%
\begin{align*}
\frac{1}{n}\mathbb{E}[W_{n}^{\mathtt{AD}}(X_{1},X_{2})^{2}]  & \leq\frac
{1}{nh_{n}^{d}}\left\{  h_{n}^{d}\mathbb{E}[K_{n}(X_{1}-X_{2})^{2}]\right\} \\
& =\frac{1}{nh_{n}^{d}}\left\{  h_{n}^{d}\int_{\mathbb{R}^{d}}\int%
_{\mathbb{R}^{d}}K_{n}(u-v)^{2}f_{0}(u)f_{0}(v)dudv\right\} \\
& =\frac{1}{nh_{n}^{d}}\int_{\mathbb{R}^{d}}\int_{\mathbb{R}^{d}}K(t)^{2}%
f_{0}(v+h_{n}t)f_{0}(v)dtdv\\
& \leq\frac{1}{nh_{n}^{d}}\left\{  \sup_{u\in\mathbb{R}^{d}}|K(u)|\right\}
\left\{  \sup_{x\in\mathbb{R}^{d}}f_{0}(x)\right\}  \int_{\mathbb{R}^{d}%
}|K(u)|du\rightarrow0.
\end{align*}
Also, because%
\[
\mathbb{E}[\{L_{n}^{\mathtt{AD}}(X)-L_{0}(X)\}^{2}]\leq4\mathbb{E}\left[
\{f_{n}^{\mathtt{AD}}(X)-f_{0}(X)\}^{2}\right]  ,
\]
a sufficient condition for $\left(  \ref{Linear term: mean square convergence}%
\right)  $ to hold is that%
\[
\mathbb{E}\left[  \{f_{n}^{\mathtt{AD}}(X)-f_{0}(X)\}^{2}\right]
\rightarrow0.
\]
As in Proposition 1(c) of \cite{Gine-Nickl_2008_PTRF}, the displayed condition
is satisfied if $h_{n}\rightarrow0.$ To summarize, each estimator satisfies
$\left(  \ref{Efficiency: Noise condition}\right)  $ under Conditions D, K,
and B$^{-}$.

The proof will be completed by giving conditions under which the estimators
satisfy $\left(  \ref{Efficiency: Bias condition}\right)  .$ As before,
suppose Conditions D and K are satisfied. In the notation introduced above,
the biases of $\hat{\theta}_{n}^{\mathtt{AD}}$ and $\hat{\theta}%
_{n}^{\mathtt{AD-LO}}$ are given by%
\[
\beta_{n}^{\mathtt{AD}}=\frac{K(0)}{nh_{n}^{d}}+\theta_{n}^{\mathtt{AD}%
}-\theta_{0}-\frac{\theta_{n}^{\mathtt{AD}}}{n}%
\]
and%
\[
\beta_{n}^{\mathtt{AD-LO}}=\theta_{n}^{\mathtt{AD}}-\theta_{0},
\]
respectively. Following \cite{Gine-Nickl_2008_Bernoulli}, we base our analysis
of the smoothing bias $\theta_{n}^{\mathtt{AD}}-\theta_{0}$ on the
representation%
\begin{align*}
\theta_{n}^{\mathtt{AD}}  & =\int_{\mathbb{R}^{d}}\int_{\mathbb{R}^{d}}%
K_{n}(u-v)f_{0}(v)f_{0}(u)dudv\\
& =\int_{\mathbb{R}^{d}}\int_{\mathbb{R}^{d}}K(t)f_{0}(u-h_{n}t)f_{0}(u)dudt\\
& =\int_{\mathbb{R}^{d}}K(t)f_{0}^{\Delta}(h_{n}t)dt,
\end{align*}
where the last equality uses the fact that $K$ is even. By Lemma 12 of
\cite{Gine-Nickl_2008_PTRF}, the function $f_{0}^{\Delta}$ belongs to the
H\"{o}lder space $\mathbf{C}^{2s}(\mathbb{R}^{d}).$ As a consequence, it
follows from standard arguments (e.g., \cite[Proposition 1.2]%
{Tsybakov_2009_Book}) that if Condition B is satisfied, then%
\[
\theta_{n}^{\mathtt{AD}}-\theta_{0}=\int_{\mathbb{R}^{d}}K(t)[f_{0}^{\Delta
}(h_{n}t)-f_{0}^{\Delta}(0)]dt=O(h_{n}^{S})=o(n^{-1/2}).
\]

In particular, $\hat{\theta}_{n}^{\mathtt{AD-LO}}$ satisfies $\left(
\ref{Efficiency: Bias condition}\right)  $ under Conditions D, K, and B. Under
the same conditions, $\theta_{n}^{\mathtt{AD}}$ is bounded, so%
\[
\sqrt{n}(\mathbb{E}[\hat{\theta}_{n}^{\mathtt{AD}}]-\theta_{0})=\frac
{K(0)}{\sqrt{nh_{n}^{2d}}}+o(1),
\]
implying in particular that Condition B must be strengthened to Condition
B$^{+}$ for $\hat{\theta}_{n}^{\mathtt{AD}}$ to satisfy $\left(
\ref{Efficiency: Bias condition}\right)  $ (unless $K(0)=0$).

Finally, the results for $\hat{\theta}_{n}^{\mathtt{AD-BC}}$ and $\hat{\theta
}_{n}^{\mathtt{AD-GJ}}$ follow from those for $\hat{\theta}_{n}^{\mathtt{AD}%
}.$ To be specific, $\hat{\theta}_{n}^{\mathtt{AD-BC}}$ differs from
$\hat{\theta}_{n}^{\mathtt{AD}}$ by an additive constant, so it satisfies
$\left(  \ref{Efficiency: Noise condition}\right)  $ under Conditions D, K,
and B$^{-}$. Also, the additive constant is designed to ensure that $\left(
\ref{Efficiency: Bias condition}\right)  $ is satisfied by $\hat{\theta}%
_{n}^{\mathtt{AD-BC}}$ under Conditions D, K, and B. Similarly, because%
\[
\frac{1}{1-c^{d}}-\frac{c^{d}}{1-c^{d}}=1,
\]
the estimator $\hat{\theta}_{n}^{\mathtt{AD-GJ}}$ satisfies $\left(
\ref{Efficiency: Noise condition}\right)  $ under Conditions D, K, and B$^{-}%
$, while the fact that%
\[
\frac{1}{1-c^{d}}\frac{1}{nh_{n}^{d}}-\frac{c^{d}}{1-c^{d}}\frac{1}%
{n(ch_{n})^{d}}=0
\]
ensures that $\left(  \ref{Efficiency: Bias condition}\right)  $ is satisfied
by $\hat{\theta}_{n}^{\mathtt{AD-GJ}}$ under Conditions D, K, and B.

\subsection{Proof of Theorem \ref{[Theorem] Bootstrap consistency}}

The estimators $\hat{\theta}_{n}^{\mathtt{AD,}\ast}$ and $\hat{\theta}%
_{n}^{\mathtt{AD-LO,}\ast}$ both have Hoeffding decompositions of the form
$\left(  \ref{Hoeffding decomposition (bootstrap)}\right)  ,$ with%
\[
L_{i,n}^{\ast}=\lambda_{i,n}\hat{L}_{n}^{\mathtt{AD}}(X_{i,n}^{\ast
})\text{\qquad and\qquad}W_{ij,n}^{\ast}=\omega_{ij,n}\hat{W}_{n}%
^{\mathtt{AD}}(X_{i,n}^{\ast},X_{j,n}^{\ast}),
\]
where $\lambda_{i,n}$ and $\omega_{ij,n}$ are the same as those for
$\hat{\theta}_{n}^{\mathtt{AD}}$ and $\hat{\theta}_{n}^{\mathtt{AD-LO}},$
while%
\[
\hat{L}_{n}^{\mathtt{AD}}(x)=2\{\hat{f}_{n}(x)-\hat{\theta}_{n}^{\mathtt{AD}%
}\},
\]%
\[
\hat{W}_{n}^{\mathtt{AD}}(x_{1},x_{2})=K_{n}(x_{1}-x_{2})-\hat{f}_{n}%
(x_{1})-\hat{f}_{n}(x_{2})+\hat{\theta}_{n}^{\mathtt{AD}}.
\]

Because the weights satisfy $\left(  \ref{Linear term: weights}\right)  $ and
$\left(  \ref{Quadratic term: weights}\right)  ,$ it follows from simple
moment calculations that the estimators satisfy%
\[
\sqrt{n}(\hat{\theta}_{n}^{\ast}-\mathbb{E}_{n}^{\ast}[\hat{\theta}_{n}^{\ast
}])=\frac{1}{\sqrt{n}}\sum_{1\leq i\leq n}\{L_{0}(X_{i,n}^{\ast}%
)-\mathbb{E}_{n}^{\ast}[L_{0}(X_{i,n}^{\ast})]\}+o_{\mathbb{P}}%
(1)\rightsquigarrow_{\mathbb{P}}\mathcal{N}(0,\sigma_{0}^{2})
\]
if%
\begin{equation}
\frac{1}{n}\mathbb{E}_{n}^{\ast}[\hat{W}_{n}^{\mathtt{AD}}(X_{1,n}^{\ast
},X_{2,n}^{\ast})^{2}]\rightarrow_{\mathbb{P}}%
0\label{Quadratic term (bootstrap): variance}%
\end{equation}
and if $\left(  \ref{Linear term: mean square convergence}\right)  $ and
$\left(  \ref{Linear term (bootstrap): mean square convergence}\right)  $
hold, where%
\begin{equation}
\mathbb{E}_{n}^{\ast}[\{\hat{L}_{n}^{\mathtt{AD}}(X_{1,n}^{\ast}%
)-L_{n}^{\mathtt{AD}}(X_{1,n}^{\ast})\}^{2}]\rightarrow_{\mathbb{P}%
}0.\label{Linear term (bootstrap): mean square convergence}%
\end{equation}

Suppose Conditions D and K are satisfied. Then $\left(
\ref{Quadratic term (bootstrap): variance}\right)  $ holds if $nh_{n}%
^{d}\rightarrow\infty,$ because then%
\begin{align*}
\frac{1}{n}\mathbb{E}_{n}^{\ast}[\hat{W}_{n}^{\mathtt{AD}}(X_{1,n}^{\ast
},X_{2,n}^{\ast})^{2}]  & \leq\frac{1}{n}\mathbb{E}_{n}^{\ast}[K_{n}%
(X_{1,n}^{\ast}-X_{2,n}^{\ast})^{2}]\\
& =\frac{1}{n^{3}}\sum_{1\leq i,j\leq n}K_{n}(X_{i}-X_{j})^{2}\\
& =\frac{1}{n^{3}}\sum_{1\leq i\leq n}K_{n}(0)^{2}+\frac{2}{n^{3}}\sum_{1\leq
i,j\leq n,i<j}K_{n}(X_{i}-X_{j})^{2}\\
& =\frac{1}{n}\left(  \frac{K(0)}{nh_{n}^{d}}\right)  ^{2}+O_{\mathbb{P}%
}\left(  \frac{1}{n}\mathbb{E}[K_{n}(X_{1}-X_{2})^{2}]\right)  \rightarrow
_{\mathbb{P}}0,
\end{align*}
where the convergence result follow from the proof of Theorem
\ref{[Theorem] Bootstrap consistency}. In that same proof it was shown that
$\left(  \ref{Linear term: mean square convergence}\right)  $ holds when
$h_{n}\rightarrow0.$ Finally, because%
\[
\mathbb{E}_{n}^{\ast}[\{\hat{L}_{n}^{\mathtt{AD}}(X_{1,n}^{\ast}%
)-L_{n}^{\mathtt{AD}}(X_{1,n}^{\ast})\}^{2}]=\frac{1}{n}\sum_{1\leq i\leq
n}\{\hat{L}_{n}^{\mathtt{AD}}(X_{i})-L_{n}^{\mathtt{AD}}(X_{i})\}^{2},
\]
a sufficient condition for $\left(
\ref{Linear term (bootstrap): mean square convergence}\right)  $ to hold is
that%
\[
\mathbb{E}[\{\hat{L}_{n}^{\mathtt{AD}}(X_{1})-L_{n}^{\mathtt{AD}}(X_{1}%
)\}^{2}]\rightarrow0.
\]
It follows from a direct calculation this condition is satisfied when
$h_{n}\rightarrow0$ and $nh_{n}^{d}\rightarrow\infty.$ To summarize, each
estimator satisfies $\left(  \ref{Bootstrap consistency: Noise condition}%
\right)  $ under Conditions D, K, and B$^{-}$.

The proof will be completed by giving conditions under which the estimators
satisfy $\left(  \ref{Bootstrap consistency: Bias condition}\right)  .$
Suppose Conditions D, K, and B are satisfied. By the proof of Theorem
\ref{[Theorem] Efficiency},%
\[
\sqrt{n}(\mathbb{E}[\hat{\theta}_{n}^{\mathtt{AD}}]-\theta_{0})=\frac
{K(0)}{\sqrt{nh_{n}^{2d}}}+o(1),
\]
and%
\[
\sqrt{n}(\mathbb{E}[\hat{\theta}_{n}^{\mathtt{AD-LO}}]-\theta_{0})=o(1),
\]
while it follows from $\left(  \ref{Hoeffding decomposition (bootstrap)}%
\right)  $ and Theorem \ref{[Theorem] Efficiency} that%
\[
\sqrt{n}(\mathbb{E}_{n}^{\ast}[\hat{\theta}_{n}^{\mathtt{AD,}\ast}%
]-\hat{\theta}_{n}^{\mathtt{AD}})=\frac{K(0)}{\sqrt{nh_{n}^{2d}}}-\frac
{\hat{\theta}_{n}^{\mathtt{AD}}}{\sqrt{n}}=\frac{K(0)}{\sqrt{nh_{n}^{2d}}%
}+o_{\mathbb{P}}(1),
\]
and%
\[
\sqrt{n}(\mathbb{E}_{n}^{\ast}[\hat{\theta}_{n}^{\mathtt{AD-LO,}\ast}%
]-\hat{\theta}_{n}^{\mathtt{AD-LO}})=\sqrt{n}(\hat{\theta}_{n}^{\mathtt{AD}%
}-\hat{\theta}_{n}^{\mathtt{AD-LO}})=\frac{K(0)}{\sqrt{nh_{n}^{2d}}%
}+o_{\mathbb{P}}(1).
\]
As a consequence, $\hat{\theta}_{n}^{\mathtt{AD,}\ast}$ satisfies $\left(
\ref{Bootstrap consistency: Bias condition}\right)  $ under Conditions D, K,
and B, whereas Condition B must be strengthened to Condition B$^{+}$ for
$\hat{\theta}_{n}^{\mathtt{AD-LO,}\ast}$ to satisfy $\left(
\ref{Bootstrap consistency: Bias condition}\right)  $ (unless $K(0)=0$).

Finally, the results for $\hat{\theta}_{n}^{\mathtt{AD-BC,}\ast}$ and
$\hat{\theta}_{n}^{\mathtt{AD-GJ,}\ast}$ follow from those for $\hat{\theta
}_{n}^{\mathtt{AD,}\ast}.$ To be specific, $\hat{\theta}_{n}^{\mathtt{AD-BC,}%
\ast} $ satisfies $\left(  \ref{Bootstrap consistency: Noise condition}%
\right)  $ under Conditions D, K, and B$^{-}$ because $\hat{\theta}%
_{n}^{\mathtt{AD,}\ast}$ does. Moreover,%
\[
\sqrt{n}(\mathbb{E}_{n}^{\ast}[\hat{\theta}_{n}^{\mathtt{AD-BC,}\ast}%
]-\hat{\theta}_{n}^{\mathtt{AD-BC}})=\sqrt{n}(\mathbb{E}_{n}^{\ast}%
[\hat{\theta}_{n}^{\mathtt{AD,}\ast}]-\hat{\theta}_{n}^{\mathtt{AD}}),
\]
so under Conditions D and K, Condition B must be strengthened to Condition
B$^{+}$ for $\hat{\theta}_{n}^{\mathtt{AD-BC,}\ast}$ to satisfy $\left(
\ref{Bootstrap consistency: Bias condition}\right)  $ (unless $K(0)=0$).
Similarly, because%
\[
\frac{1}{1-c^{d}}-\frac{c^{d}}{1-c^{d}}=1,
\]
the estimator $\hat{\theta}_{n}^{\mathtt{AD-GJ,}\ast}$ satisfies $\left(
\ref{Bootstrap consistency: Noise condition}\right)  $ under Conditions D, K,
and B$^{-}$, while the fact that%
\[
\frac{1}{1-c^{d}}\frac{1}{nh_{n}^{d}}-\frac{c^{d}}{1-c^{d}}\frac{1}%
{n(ch_{n})^{d}}=0
\]
ensures that $\left(  \ref{Bootstrap consistency: Bias condition}\right)  $ is
satisfied by $\hat{\theta}_{n}^{\mathtt{AD-GJ,}\ast}$ under Conditions D, K,
and B.

\subsection{Proof of Theorem \ref{[Theorem] Efficiency (ISD)}}

The proof is similar to that of Theorem \ref{[Theorem] Efficiency}. The
estimators $\hat{\theta}_{n}^{\mathtt{ISD}},\ \hat{\theta}_{n}%
^{\mathtt{ISD-LO}},$ and $\hat{\theta}_{n}^{\mathtt{ISD-CF}}$ all have
Hoeffding decompositions of the form $\left(  \ref{Hoeffding decomposition}%
\right)  ,$ with%
\[
L_{i,n}=\lambda_{i,n}L_{n}^{\mathtt{ISD}}(X_{i}),\text{\qquad}W_{ij,n}%
=\omega_{ij,n}W_{n}^{\mathtt{ISD}}(X_{i},X_{j}),
\]
where $\lambda_{i,n}$ and $\omega_{ij,n}$ are (non-random) estimator-specific
weights, while%
\[
L_{n}^{\mathtt{ISD}}(x)=2\{f_{n}^{\mathtt{ISD}}(x)-\theta_{n}^{\mathtt{ISD}%
}\},
\]%
\[
W_{n}^{\mathtt{ISD}}(x_{1},x_{2})=K_{n}^{\Delta}(x_{1}-x_{2})-f_{n}%
^{\mathtt{ISD}}(x_{1})-f_{n}^{\mathtt{ISD}}(x_{2})+\theta_{n}^{\mathtt{ISD}},
\]
where%

\[
f_{n}^{\mathtt{ISD}}\left(  x\right)  =\mathbb{E}[K_{n}^{\Delta}%
(x-X)]=\int_{\mathbb{R}^{d}}K^{\Delta}(u)f_{0}(x+uh_{n})du,
\]%
\[
\theta_{n}^{\mathtt{ISD}}=\mathbb{E}[f_{n}^{\mathtt{ISD}}(X)]=\int%
_{\mathbb{R}^{d}}f_{n}^{\mathtt{ISD}}(x)f_{0}(x)dx,
\]%
\[
K_{n}^{\Delta}(x)=\frac{1}{h_{n}^{d}}K^{\Delta}\left(  \frac{x}{h_{n}}\right)
,\qquad K^{\Delta}(x)=\int_{\mathbb{R}^{d}}K(u)K(x+u)du.
\]

To be specific, in the case of%
\begin{align*}
\hat{\theta}_{n}^{\mathtt{ISD}}  & =\int_{\mathbb{R}^{d}}\hat{f}_{n}\left(
x\right)  ^{2}dx\\
& =\int_{\mathbb{R}^{d}}\left[  \frac{1}{n}\sum_{1\leq j_{1}\leq n}%
K_{n}(x-X_{j_{1}})\right]  \left[  \frac{1}{n}\sum_{1\leq j_{2}\leq n}%
K_{n}(x-X_{j_{2}})\right]  dx\\
& =\frac{1}{n^{2}}\sum_{1\leq i,j\leq n}K_{n}^{\Delta}(X_{i}-X_{j}),
\end{align*}
each $\lambda_{i,n}$ and $\omega_{ij,n}$ is given by $1-n^{-1}.$ For%
\begin{align*}
\hat{\theta}_{n}^{\mathtt{ISD-LO}}  & =\frac{1}{n}\sum_{1\leq i\leq n}%
\int_{\mathbb{R}^{d}}\hat{f}_{i,n}^{\mathtt{LO}}(x)^{2}dx\\
& =\frac{1}{n}\sum_{1\leq i\leq n}\int_{\mathbb{R}^{d}}\left[  \sum_{1\leq
j_{1}\leq n}w_{ij_{1},n}K_{n}(x-X_{j_{1}})\right]  \left[  \sum_{1\leq
j_{2}\leq n}w_{ij_{2},n}K_{n}(x-X_{j_{2}})\right]  dx\\
& =\frac{1}{n^{2}}\sum_{1\leq i,j\leq n}\left[  n\sum_{1\leq k\leq n}%
w_{ki,n}w_{kj,n}\right]  K_{n}^{\Delta}(X_{i}-X_{j}),
\end{align*}
the weights are given by%
\[
\lambda_{i,n}=\sum_{1\leq j,k\leq n,j\neq i}w_{ki,n}w_{kj,n},\qquad
\omega_{ij,n}=(n-1)\sum_{1\leq k\leq n}w_{ki,n}w_{kj,n},
\]
while the weights for%
\begin{align*}
\hat{\theta}_{n}^{\mathtt{ISD-DCF}}  & =\int_{\mathbb{R}^{d}}\hat{f}%
_{1,n}^{\mathtt{CF}}(x)\hat{f}_{n,n}^{\mathtt{CF}}(x)dx\\
& =\int_{\mathbb{R}^{d}}\left[  \sum_{1\leq j_{1}\leq n}w_{1j_{1},n}%
K_{n}(x-X_{j_{1}})\right]  \left[  \sum_{1\leq j_{2}\leq n}w_{nj_{2},n}%
K_{n}(x-X_{j_{2}})\right]  dx\\
& =\frac{1}{n^{2}}\sum_{1\leq i,j\leq n}[n^{2}w_{1i,n}w_{nj,n}]K_{n}^{\Delta
}(X_{i}-X_{j})
\end{align*}
can be shown to be given by%

\[
\lambda_{i,n}=\frac{n/2}{\sum_{1\leq j\leq n}%
\I
(\left\lceil 2i/n\right\rceil =\left\lceil 2j/n\right\rceil )},\qquad
\omega_{ij,n}^{\mathtt{ISD-DCF}}=\frac{n(n-1)/2}{(n-\left\lfloor
n/2\right\rfloor )\left\lfloor n/2\right\rfloor }%
\I
(\left\lceil 2i/n\right\rceil \neq\left\lceil 2j/n\right\rceil ).
\]

In all cases, the weights satisfy $\left(  \ref{Linear term: weights}\right)
$ and $\left(  \ref{Quadratic term: weights}\right)  ,$ so the estimators
satisfy $\left(  \ref{Efficiency: Noise condition}\right)  $ if%
\begin{equation}
\frac{1}{n}\mathbb{E}[W_{n}^{\mathtt{ISD}}(X_{1},X_{2})^{2}]\rightarrow
0\label{Quadratic term: variance (ISD)}%
\end{equation}
and if%
\begin{equation}
\mathbb{E}[\{L_{n}^{\mathtt{ISD}}(X)-L_{0}(X)\}^{2}]\rightarrow
0.\label{Linear term: mean square convergence (ISD)}%
\end{equation}
Proceeding as in the proof of Theorem \ref{[Theorem] Efficiency} it can be
shown that $\left(  \ref{Quadratic term: variance (ISD)}\right)  $ and
$\left(  \ref{Linear term: mean square convergence (ISD)}\right)  $ hold under
Conditions D, K, and B$^{-}$.

Finally, the biases of $\hat{\theta}_{n}^{\mathtt{ISD}},\ \hat{\theta}%
_{n}^{\mathtt{ISD-LO}},$ and $\hat{\theta}_{n}^{\mathtt{ISD-DCF}}$ are given
by%
\[
\beta_{n}^{\mathtt{ISD}}=\frac{K^{\Delta}(0)}{nh_{n}^{d}}+\theta
_{n}^{\mathtt{ISD}}-\theta_{0}-\frac{\theta_{n}^{\mathtt{ISD}}}{n},
\]%
\[
\beta_{n}^{\mathtt{ISD-LO}}=\eta_{n}\frac{K^{\Delta}(0)}{nh_{n}^{d}}%
+\theta_{n}^{\mathtt{ISD}}-\theta_{0}-\eta_{n}\frac{\theta_{n}^{\mathtt{ISD}}%
}{n},
\]
and%
\[
\beta_{n}^{\mathtt{ISD-DCF}}=\theta_{n}^{\mathtt{ISD}}-\theta_{0},
\]
respectively, where%
\[
\eta_{n}=\sum_{1\leq i\leq n}\frac{1}{\sum_{1\leq j\leq n}%
\I
(\left\lceil iB_{n}/n\right\rceil \neq\left\lceil jB_{n}/n\right\rceil )},
\]
and where%
\[
\theta_{n}^{\mathtt{ISD}}-\theta_{0}=\int_{\mathbb{R}^{d}}K^{\Delta}%
(t)[f_{0}^{\Delta}(h_{n}t)-f_{0}^{\Delta}(0)]dt=O(h_{n}^{S})=o(n^{-1/2})
\]
under Conditions D, K, and B.

As a consequence $\hat{\theta}_{n}^{\mathtt{ISD-DCF}}$ satisfies $\left(
\ref{Efficiency: Bias condition}\right)  $ under Conditions D, K, and B,
whereas%
\[
\sqrt{n}(\mathbb{E}[\hat{\theta}_{n}^{\mathtt{ISD}}]-\theta_{0})=\frac
{K^{\Delta}(0)}{\sqrt{nh_{n}^{2d}}}+o(1),
\]
so Condition B must be strengthened to Condition B$^{+}$ for $\hat{\theta}%
_{n}^{\mathtt{ISD}}$ to satisfy $\left(  \ref{Efficiency: Bias condition}%
\right)  .$ Finally, $\eta_{n}\geq1$ is bounded, so Condition B must be
strengthened to Condition B$^{+}$ for $\hat{\theta}_{n}^{\mathtt{ISD-LO}}$ to
satisfy $\left(  \ref{Efficiency: Bias condition}\right)  .$

Finally, the results for $\hat{\theta}_{n}^{\mathtt{ISD-BC}}$ and $\hat
{\theta}_{n}^{\mathtt{ISD-GJ}}$ follow from those for $\hat{\theta}%
_{n}^{\mathtt{ISD}}.$ To be specific, $\hat{\theta}_{n}^{\mathtt{ISD-BC}}$
differs from $\hat{\theta}_{n}^{\mathtt{ISD}}$ by an additive constant, so it
satisfies $\left(  \ref{Efficiency: Noise condition}\right)  $ under
Conditions D, K, and B$^{-}$. Also, the additive constant is designed to
ensure that $\left(  \ref{Efficiency: Bias condition}\right)  $ is satisfied
by $\hat{\theta}_{n}^{\mathtt{ISD-BC}}$ under Conditions D, K, and B.
Similarly, because%
\[
\frac{1}{1-c^{d}}-\frac{c^{d}}{1-c^{d}}=1,
\]
the estimator $\hat{\theta}_{n}^{\mathtt{ISD-GJ}}$ satisfies $\left(
\ref{Efficiency: Noise condition}\right)  $ under Conditions D, K, and B$^{-}%
$, while the fact that%
\[
\frac{1}{1-c^{d}}\frac{1}{nh_{n}^{d}}-\frac{c^{d}}{1-c^{d}}\frac{1}%
{n(ch_{n})^{d}}=0
\]
ensures that $\left(  \ref{Efficiency: Bias condition}\right)  $ is satisfied
by $\hat{\theta}_{n}^{\mathtt{ISD-GJ}}$ under Conditions D, K, and B.

\subsection{Proof of Theorem \ref{[Theorem] Bootstrap consistency (ISD)}}

The proof is similar to that of Theorem \ref{[Theorem] Bootstrap consistency}.
The estimators $\hat{\theta}_{n}^{\mathtt{ISD,}\ast},\ \hat{\theta}%
_{n}^{\mathtt{ISD-LO,}\ast},$ and $\hat{\theta}_{n}^{\mathtt{ISD-DCF,}\ast}$
all have Hoeffding decompositions of the form $\left(
\ref{Hoeffding decomposition (bootstrap)}\right)  ,$ with%
\[
L_{i,n}^{\ast}=\lambda_{i,n}\hat{L}_{n}^{\mathtt{ISD}}(X_{i,n}^{\ast
}),\text{\qquad}W_{ij,n}^{\ast}=\omega_{ij,n}\hat{W}_{n}^{\mathtt{ISD}%
}(X_{i,n}^{\ast},X_{j,n}^{\ast}),
\]
where $\lambda_{i,n}$ and $\omega_{ij,n}$ are the same as those for
$\hat{\theta}_{n}^{\mathtt{ISD}},\ \hat{\theta}_{n}^{\mathtt{ISD-LO}},$ and
$\hat{\theta}_{n}^{\mathtt{ISD-DCF}},$ while%
\[
\hat{L}_{n}^{\mathtt{ISD}}(x)=2\{\hat{f}_{n}^{\mathtt{ISD}}(x)-\hat{\theta
}_{n}^{\mathtt{ISD}}\},\qquad\hat{f}_{n}^{\mathtt{ISD}}(x)=\frac{1}{n}%
\sum_{1\leq j\leq n}K_{n}^{\Delta}(x-X_{j}),
\]%
\[
\hat{W}_{n}^{\mathtt{ISD}}(x_{1},x_{2})=K_{n}^{\Delta}(x_{1}-x_{2})-\hat
{f}_{n}^{\mathtt{ISD}}(x_{1})-\hat{f}_{n}^{\mathtt{ISD}}(x_{2})+\hat{\theta
}_{n}^{\mathtt{ISD}}.
\]

Because the weights satisfy $\left(  \ref{Linear term: weights}\right)  $ and
$\left(  \ref{Quadratic term: weights}\right)  ,$ it follows from simple
moment calculations that the estimators satisfy%
\[
\sqrt{n}(\hat{\theta}_{n}^{\ast}-\mathbb{E}_{n}^{\ast}[\hat{\theta}_{n}^{\ast
}])=\frac{1}{\sqrt{n}}\sum_{1\leq i\leq n}\{L_{0}(X_{i,n}^{\ast}%
)-\mathbb{E}_{n}^{\ast}[L_{0}(X_{i,n}^{\ast})]\}+o_{\mathbb{P}}%
(1)\rightsquigarrow_{\mathbb{P}}\mathcal{N}(0,\sigma_{0}^{2})
\]
if%
\begin{equation}
\frac{1}{n}\mathbb{E}_{n}^{\ast}[\hat{W}_{n}^{\mathtt{ISD}}(X_{1,n}^{\ast
},X_{2,n}^{\ast})^{2}]\rightarrow_{\mathbb{P}}%
0\label{Quadratic term (bootstrap): variance (ISD)}%
\end{equation}
and if $\left(  \ref{Linear term: mean square convergence (ISD)}\right)  $ and
$\left(  \ref{Linear term (bootstrap): mean square convergence (ISD)}\right)
$ hold, where%
\begin{equation}
\mathbb{E}_{n}^{\ast}[\{\hat{L}_{n}^{\mathtt{ISD}}(X_{1,n}^{\ast}%
)-L_{n}^{\mathtt{ISD}}(X_{1,n}^{\ast})\}^{2}]\rightarrow_{\mathbb{P}%
}0.\label{Linear term (bootstrap): mean square convergence (ISD)}%
\end{equation}

Suppose Conditions D and K are satisfied. Then $\left(
\ref{Quadratic term (bootstrap): variance (ISD)}\right)  $ holds if
$nh_{n}^{d}\rightarrow\infty,$ because then%
\begin{align*}
\frac{1}{n}\mathbb{E}_{n}^{\ast}[\hat{W}_{n}^{\mathtt{ISD}}(X_{1,n}^{\ast
},X_{2,n}^{\ast})^{2}]  & \leq\frac{1}{n}\mathbb{E}_{n}^{\ast}[K_{n}^{\Delta
}(X_{1,n}^{\ast}-X_{2,n}^{\ast})^{2}]\\
& =\frac{1}{n^{3}}\sum_{1\leq i,j\leq n}K_{n}^{\Delta}(X_{i}-X_{j})^{2}\\
& =\frac{1}{n^{3}}\sum_{1\leq i\leq n}K_{n}^{\Delta}(0)^{2}+\frac{2}{n^{3}%
}\sum_{1\leq i,j\leq n,i<j}K_{n}^{\Delta}(X_{i}-X_{j})^{2}\\
& =\frac{1}{n}\left(  \frac{K^{\Delta}(0)}{nh_{n}^{d}}\right)  ^{2}%
+O_{\mathbb{P}}\left(  \frac{1}{n}\mathbb{E}[K_{n}^{\Delta}(X_{1}-X_{2}%
)^{2}]\right)  \rightarrow_{\mathbb{P}}0.
\end{align*}
Also, $\left(  \ref{Linear term: mean square convergence (ISD)}\right)  $
holds when $h_{n}\rightarrow0.$ Finally, because%
\[
\mathbb{E}_{n}^{\ast}[\{\hat{L}_{n}^{\mathtt{ISD}}(X_{1,n}^{\ast}%
)-L_{n}^{\mathtt{ISD}}(X_{1,n}^{\ast})\}^{2}]=\frac{1}{n}\sum_{1\leq i\leq
n}\{\hat{L}_{n}^{\mathtt{ISD}}(X_{i})-L_{n}^{\mathtt{ISD}}(X_{i})\}^{2},
\]
a sufficient condition for $\left(
\ref{Linear term (bootstrap): mean square convergence (ISD)}\right)  $ to hold
is that%
\[
\mathbb{E}[\{\hat{L}_{n}^{\mathtt{ISD}}(X_{1})-L_{n}^{\mathtt{ISD}}%
(X_{1})\}^{2}]\rightarrow0.
\]
It follows from a direct calculation this condition is satisfied when
$h_{n}\rightarrow0$ and $nh_{n}^{d}\rightarrow\infty.$ To summarize, each
estimator satisfies $\left(  \ref{Bootstrap consistency: Noise condition}%
\right)  $ under Conditions D, K, and B$^{-}$.

The proof will be completed by giving conditions under which the estimators
satisfy $\left(  \ref{Bootstrap consistency: Bias condition}\right)  .$
Suppose Conditions D, K, and B are satisfied. By the proof of Theorem
\ref{[Theorem] Efficiency (ISD)},%
\[
\sqrt{n}(\mathbb{E}[\hat{\theta}_{n}^{\mathtt{ISD}}]-\theta_{0})=\frac
{K^{\Delta}(0)}{\sqrt{nh_{n}^{2d}}}+o(1),
\]%
\[
\sqrt{n}(\mathbb{E}[\hat{\theta}_{n}^{\mathtt{ISD-LO}}]-\theta_{0})=\eta
_{n}\frac{K^{\Delta}(0)}{\sqrt{nh_{n}^{2d}}}+o(1),
\]
and%
\[
\sqrt{n}(\mathbb{E}[\hat{\theta}_{n}^{\mathtt{ISD-DCF}}]-\theta_{0})=o(1),
\]
while it follows from $\left(  \ref{Hoeffding decomposition (bootstrap)}%
\right)  $ and Theorem \ref{[Theorem] Efficiency (ISD)} that%
\[
\sqrt{n}(\mathbb{E}_{n}^{\ast}[\hat{\theta}_{n}^{\mathtt{ISD,}\ast}%
]-\hat{\theta}_{n}^{\mathtt{ISD}})=\frac{K^{\Delta}(0)}{\sqrt{nh_{n}^{2d}}%
}-\frac{\hat{\theta}_{n}^{\mathtt{ISD}}}{\sqrt{n}}=\frac{K^{\Delta}(0)}%
{\sqrt{nh_{n}^{2d}}}+o_{\mathbb{P}}(1),
\]%
\[
\sqrt{n}(\mathbb{E}_{n}^{\ast}[\hat{\theta}_{n}^{\mathtt{ISD-LO,}\ast}%
]-\hat{\theta}_{n}^{\mathtt{ISD-LO}})=\eta_{n}\frac{K^{\Delta}(0)}%
{\sqrt{nh_{n}^{2d}}}+\sqrt{n}(\hat{\theta}_{n}^{\mathtt{ISD}}-\hat{\theta}%
_{n}^{\mathtt{ISD-LO}})-\eta_{n}\frac{\hat{\theta}_{n}^{\mathtt{ISD}}}%
{\sqrt{n}}=\frac{K^{\Delta}(0)}{\sqrt{nh_{n}^{2d}}}+o_{\mathbb{P}}(1).
\]
and%
\[
\sqrt{n}(\mathbb{E}_{n}^{\ast}[\hat{\theta}_{n}^{\mathtt{ISD-DCF,}\ast}%
]-\hat{\theta}_{n}^{\mathtt{ISD-DCF}})=\sqrt{n}(\hat{\theta}_{n}%
^{\mathtt{ISD}}-\hat{\theta}_{n}^{\mathtt{ISD-DCF}})=\frac{K^{\Delta}%
(0)}{\sqrt{nh_{n}^{2d}}}+o_{\mathbb{P}}(1).
\]
As a consequence, $\hat{\theta}_{n}^{\mathtt{ISD,}\ast}$ satisfies $\left(
\ref{Bootstrap consistency: Bias condition}\right)  $ under Conditions D, K,
and B, whereas Condition B must be strengthened to Condition B$^{+}$ for
$\hat{\theta}_{n}^{\mathtt{ISD-DCF,}\ast}$ to satisfy $\left(
\ref{Bootstrap consistency: Bias condition}\right)  .$ Finally, if $B_{n}=n,$
then%
\[
\eta_{n}=\frac{n}{n-1}=1+O(n^{-1}),
\]
so $\hat{\theta}_{n}^{\mathtt{ISD-LO,}\ast}$ satisfies $\left(
\ref{Bootstrap consistency: Bias condition}\right)  $ under Conditions D, K,
and B. Under the other hand, Condition B must be strengthened to Condition
B$^{+} $ for the cross-fit version of $\hat{\theta}_{n}^{\mathtt{ISD-LO,}\ast}
$ to satisfy $\left(  \ref{Bootstrap consistency: Bias condition}\right)  $
because if $B_{n}=B$ for all $n,$ then%
\[
\eta_{n}\rightarrow\frac{B}{B-1}\neq1.
\]

Finally, the results for $\hat{\theta}_{n}^{\mathtt{ISD-BC,}\ast}$ and
$\hat{\theta}_{n}^{\mathtt{ISD-GJ,}\ast}$ follow from those for $\hat{\theta
}_{n}^{\mathtt{ISD,}\ast}.$ To be specific, $\hat{\theta}_{n}%
^{\mathtt{ISD-BC,}\ast}$ satisfies $\left(
\ref{Bootstrap consistency: Noise condition}\right)  $ under Conditions D, K,
and B$^{-}$ because $\hat{\theta}_{n}^{\mathtt{ISD,}\ast}$ does. Moreover,%
\[
\sqrt{n}(\mathbb{E}_{n}^{\ast}[\hat{\theta}_{n}^{\mathtt{ISD-BC,}\ast}%
]-\hat{\theta}_{n}^{\mathtt{ISD-BC}})=\sqrt{n}(\mathbb{E}_{n}^{\ast}%
[\hat{\theta}_{n}^{\mathtt{ISD,}\ast}]-\hat{\theta}_{n}^{\mathtt{ISD}}),
\]
so under Conditions D and K, Condition B must be strengthened to Condition
B$^{+}$ for $\hat{\theta}_{n}^{\mathtt{ISD-BC,}\ast}$ to satisfy $\left(
\ref{Bootstrap consistency: Bias condition}\right)  $ (unless $K(0)=0$).
Similarly, because%
\[
\frac{1}{1-c^{d}}-\frac{c^{d}}{1-c^{d}}=1,
\]
the estimator $\hat{\theta}_{n}^{\mathtt{ISD-GJ,}\ast}$ satisfies $\left(
\ref{Bootstrap consistency: Noise condition}\right)  $ under Conditions D, K,
and B$^{-}$, while the fact that%
\[
\frac{1}{1-c^{d}}\frac{1}{nh_{n}^{d}}-\frac{c^{d}}{1-c^{d}}\frac{1}%
{n(ch_{n})^{d}}=0
\]
ensures that $\left(  \ref{Bootstrap consistency: Bias condition}\right)  $ is
satisfied by $\hat{\theta}_{n}^{\mathtt{ISD-GJ,}\ast}$ under Conditions D, K,
and B.

\subsection{Proof of Theorem \ref{[Theorem] Efficiency (LR)}}

It follows from the proofs of Theorems \ref{[Theorem] Efficiency} and
\ref{[Theorem] Efficiency (ISD)} that the estimators $\hat{\theta}%
_{n}^{\mathtt{LR}},$ $\hat{\theta}_{n}^{\mathtt{LR-BC}},$ $\hat{\theta}%
_{n}^{\mathtt{LR-GJ}},$ and $\hat{\theta}_{n}^{\mathtt{LR-LO}}$ satisfy
$\left(  \ref{Efficiency: Noise condition}\right)  $ under Conditions D, K,
and B$^{-}$ and have biases of the form%
\[
\beta_{n}^{\mathtt{LR}}=2\beta_{n}^{\mathtt{AD}}-\beta_{n}^{\mathtt{ISD}%
}=\frac{2K(0)-K^{\Delta}(0)}{nh_{n}^{d}}+o(n^{-1/2}),
\]%
\[
\beta_{n}^{\mathtt{LR-BC}}=o(n^{-1/2}),\qquad\beta_{n}^{\mathtt{LR-GJ}%
}=o(n^{-1/2}),
\]
and%
\[
\beta_{n}^{\mathtt{LR-LO}}=2\beta_{n}^{\mathtt{AD-LO}}-\beta_{n}%
^{\mathtt{ISD-LO}}=-\eta_{n}\frac{K^{\Delta}(0)}{nh_{n}^{d}}+o(n^{-1/2}),
\]
respectively, under Conditions D, K, and B.

As a consequence $\hat{\theta}_{n}^{\mathtt{LR-BC}}$ and $\hat{\theta}%
_{n}^{\mathtt{LR-GJ}}$ satisfy $\left(  \ref{Efficiency: Bias condition}%
\right)  $ under Conditions D, K, and B, whereas Condition B must be
strengthened to Condition B$^{+}$ for $\hat{\theta}_{n}^{\mathtt{LR-LO}}$ to
satisfy $\left(  \ref{Efficiency: Bias condition}\right)  .$ Likewise,
Condition B must be strengthened to Condition B$^{+}$ for $\hat{\theta}%
_{n}^{\mathtt{LR}}$ to satisfy $\left(  \ref{Efficiency: Bias condition}%
\right)  $ unless $2K(0)=K^{\Delta}(0).$

\subsection{Proof of Theorem \ref{[Theorem] Bootstrap consistency (LR)}}

It follows from the proofs of Theorems \ref{[Theorem] Bootstrap consistency}
and \ref{[Theorem] Bootstrap consistency (ISD)} that the estimators
$\hat{\theta}_{n}^{\mathtt{LR,}\ast},$ $\hat{\theta}_{n}^{\mathtt{LR-BC,}\ast
},$ $\hat{\theta}_{n}^{\mathtt{LR-GJ,}\ast},$ and $\hat{\theta}_{n}%
^{\mathtt{LR-LO,}\ast}$ satisfy $\left(
\ref{Bootstrap consistency: Noise condition}\right)  $ under Conditions D, K,
and B$^{-}.$

The proof will be completed by giving conditions under which the estimators
satisfy $\left(  \ref{Bootstrap consistency: Bias condition}\right)  .$
Suppose Conditions D, K, and B are satisfied. By the proof of Theorem
\ref{[Theorem] Efficiency (LR)},%
\[
\sqrt{n}(\mathbb{E}[\hat{\theta}_{n}^{\mathtt{LR}}]-\theta_{0})=\frac
{2K(0)-K^{\Delta}(0)}{\sqrt{nh_{n}^{2d}}}+o(1),
\]%
\[
\sqrt{n}(\mathbb{E}[\hat{\theta}_{n}^{\mathtt{LR-BC}}]-\theta_{0})=o(1),
\]%
\[
\sqrt{n}(\mathbb{E}[\hat{\theta}_{n}^{\mathtt{LR-GJ}}]-\theta_{0})=o(1),
\]
and%
\[
\sqrt{n}(\mathbb{E}[\hat{\theta}_{n}^{\mathtt{LR-LO}}]-\theta_{0})=-\eta
_{n}\frac{K^{\Delta}(0)}{nh_{n}^{d}}+o(1),
\]
while it follows from the proofs of Theorems
\ref{[Theorem] Bootstrap consistency} and
\ref{[Theorem] Bootstrap consistency (ISD)} that%
\[
\sqrt{n}(\mathbb{E}_{n}^{\ast}[\hat{\theta}_{n}^{\mathtt{LR,}\ast}%
]-\hat{\theta}_{n}^{\mathtt{LR}})=\frac{2K(0)-K^{\Delta}(0)}{\sqrt{nh_{n}%
^{2d}}}+o_{\mathbb{P}}(1),
\]%
\[
\sqrt{n}(\mathbb{E}_{n}^{\ast}[\hat{\theta}_{n}^{\mathtt{LR-BC,}\ast}%
]-\hat{\theta}_{n}^{\mathtt{LR-BC}})=\frac{2K(0)-K^{\Delta}(0)}{\sqrt
{nh_{n}^{2d}}}+o_{\mathbb{P}}(1),
\]%
\[
\sqrt{n}(\mathbb{E}_{n}^{\ast}[\hat{\theta}_{n}^{\mathtt{LR-GJ,}\ast}%
]-\hat{\theta}_{n}^{\mathtt{LR-GJ}})=o_{\mathbb{P}}(1),
\]
and%
\[
\sqrt{n}(\mathbb{E}_{n}^{\ast}[\hat{\theta}_{n}^{\mathtt{LR-LO,}\ast}%
]-\hat{\theta}_{n}^{\mathtt{LR-LO}})=\frac{2K(0)-K^{\Delta}(0)}{\sqrt
{nh_{n}^{2d}}}+o_{\mathbb{P}}(1).
\]

As a consequence, $\hat{\theta}_{n}^{\mathtt{LR,}\ast}$ and $\hat{\theta}%
_{n}^{\mathtt{LR-GJ,}\ast}$ satisfy $\left(
\ref{Bootstrap consistency: Bias condition}\right)  $ under Conditions D, K,
and B, whereas Condition B must be strengthened to Condition B$^{+}$ for
$\hat{\theta}_{n}^{\mathtt{LR-BC,}\ast}$ and $\hat{\theta}_{n}%
^{\mathtt{LR-LO,}\ast}$ to satisfy $\left(
\ref{Bootstrap consistency: Bias condition}\right)  .$

\bibliographystyle{econometrica}
\bibliography{Cattaneo-Jansson_2020_AverageDensityEstimators.bib}

\end{document}